\pdfoutput=1
\documentclass[11pt,a4paper]{article} 
\usepackage{slashed,jheppub,multirow,xspace}
\usepackage[export]{adjustbox}
\usepackage[utf8]{inputenc}

\newcommand\sss{\scriptscriptstyle}
\newcommand{\be}{\begin{equation}}
\newcommand{\ee}{\end{equation}}
\newcommand{\beq}{\begin{eqnarray}}
\newcommand{\eeq}{\end{eqnarray}}
\newcommand{\de}{{\sc Delphes}~3}
\newcommand{\amc}{{\sc MG5\textunderscore}a{\sc MC@NLO}}
\newcommand{\fr}{{\sc Feyn\-Rules}}
\newcommand{\fa}{{\sc Feyn\-Arts}}
\newcommand{\ml}{{\sc MadLoop}}
\newcommand{\mfks}{{\sc MadFKS}}
\newcommand{\msp}{{\sc MadSpin}}
\newcommand{\mw}{{\sc MadWidth}}
\newcommand{\nloct}{{\sc NLOCT}}
\newcommand{\py}{{\sc Pythia}~8}
\newcommand{\fj}{{\sc FastJet}}
\newcommand{\Hbb}{\ensuremath{h \to b\overline{b}}\xspace}
\newcommand{\Htautau}{\ensuremath{h \to \tau^{+}\tau^{-}}\xspace}
\newcommand{\Hgamgam}{\ensuremath{h \to \gamma\gamma}\xspace}
\newcommand{\HWW}{\ensuremath{h \to W^{+}W^{-}}\xspace}
\newcommand{\pt}{\ensuremath{p_{T}}\xspace}
\def\bsp#1\esp{\begin{split}#1\end{split}}

\title{Probing vector-like quark models with Higgs-boson pair production}

\author[1]{Giacomo Cacciapaglia}
\author[1]{\!\!, Haiying Cai}
\author[2]{\!\!, Alexandra Carvalho}
\author[1,3]{\!\!, Aldo Deandrea}
\author[4]{\!\!, Thomas Flacke}
\author[3,5,6]{\!\!, Benjamin Fuks}
\author[7]{\!\!, Devdatta Majumder}
\author[5,6]{\! and Hua-Sheng Shao}

\affiliation[1]{Univ Lyon, Universit\'e Lyon 1, CNRS/IN2P3, IPNL, F-69622,
   Villeurbanne, France}
\affiliation[2]{Dipartimento di Fisica e Astronomia and INFN, Sezione di Padova,
   Via Marzolo 8, I-35131 Padova, Italy}
\affiliation[3]{Institut Universitaire de France, 103 boulevard Saint-Michel,
   75005 Paris, France}
\affiliation[4]{Center for Theoretical Physics of the Universe, Institute for
   Basic Science (IBS), Daejeon 34051 Korea}
\affiliation[5]{Sorbonne Universit\'es, UPMC Univ.~Paris 06, UMR 7589, LPTHE,
   F-75005 Paris, France}
\affiliation[6]{CNRS, UMR 7589, LPTHE, F-75005 Paris, France}
\affiliation[7]{Department of Physics and Astronomy, University of Kansas, 
1082 Malott, 1251 Wescoe Hall Dr., Lawrence, KS 66045, USA}

\abstract{
We investigate Higgs-boson pair production at the LHC when the final state
system arises from decays of vector-like quarks coupling to the
Higgs boson and the Standard Model quarks. Our phenomenological study includes
next-to-leading-order QCD corrections, which are important to guarantee accurate
predictions, and focuses on a detailed analysis of a di-Higgs signal in the four
$b$-jet channel. Whereas existing Run~II CMS and ATLAS analyses are not
specifically designed for probing non-resonant, vector-like-quark induced,
di-Higgs production, we show that they nevertheless offer some potential for
these modes. We then investigate the possibility of distinguishing between the
various di-Higgs production mechanisms by exploiting the kinematic
properties of the signal.}

\begin{document}

\hspace*{100mm}{\large \tt CTPU-17-05, LYCEN 2017-02} \\

\maketitle
\flushbottom

\section{Introduction} \label{sec:intro}

Vector-like quarks (VLQ), or quarks whose left-handed and right-handed
components lie in the same representation of the Standard Model (SM) gauge
symmetry group, are under strong scrutiny at the LHC due to their relevance in
many extensions of the SM. They appear, for instance, in models featuring
extra space-time dimensions, an extended gauge symmetry, or a composite Higgs 
sector~\cite{Antoniadis:1990ew,Kaplan:1991dc,ArkaniHamed:2002qx,%
Contino:2006qr,Matsedonskyi:2012ym}. VLQs in general mix with all three
generations of SM quarks, and the mixing with the first generation can in
particular be quite relevant for various new physics production mechanisms, even
though such a mixing is constrained by low-energy experimental data. On the
other hand, the structure of the theories containing vector-like quarks enables
a wealth of options for describing the decays of these quarks into the SM
particles, and the channel in which the VLQ decays into a Higgs boson often
plays an important role. For instance, if the model particle content exhibits
more than one VLQ multiplet that largely mixes with the SM quark sector, an
almost exclusive decay into a Higgs boson (and not any other gauge boson) and an
accompanying SM quark is
possible~\cite{Atre:2013ap,Azatov:2012rj}. Such a possibility is 
especially realised in specific and well motivated new physics models containing
two doublets of VLQs.

We focus in this work on the production of two Higgs bosons, or di-Higgs
production, in the context of models featuring VLQs that always decay into a
Higgs boson and a SM quark. Run~II LHC searches for new physics in di-Higgs
events have been performed by both the ATLAS and CMS collaborations, using
final-state signatures made of four jets issued from the
fragmentation of bottom quarks, or $b$-jets, in particular when the pair of
Higgs bosons could possibly originate from the decay of a heavier
resonance~\cite{ATLAS:2016ixk,Aaboud:2016xco,CMS:2016foy,CMS:2016tlj,%
CMS:2016pwo}. Other signatures have also been considered, when, for instance, a
resonantly produced di-Higgs system decays into a pair of $b$-jets and either a
pair of taus~\cite{CMS:2016knm,CMS:2016ymn}, a pair of photons~\cite{%
ATLAS:2016lala,CMS:2016vpz}, or a pair of weak bosons~\cite{CMS:2016cdj}, as
well as into a pair of weak bosons and a diphoton system~\cite{ATLAS:2016qmt}.

As the four-$b$-jet signature is associated with the largest branching ratio, we
investigate the potential impact of the presence of VLQs on the searches by a
study of
di-Higgs production and decay into a final state containing four $b$-jets. 
The present study is meant to give an example of the potential of the di-Higgs 
mode for investigating physics beyond the SM, rather than being a complete analysis, 
which can only be done at the level of the LHC collaborations. Extra
quarks can enhance the cross section~\cite{Brooijmans:2016vro,Fuks:2016ftf} related to the
production of two Higgs bosons, the latter being produced either directly or
through the VLQ decays. As a consequence, the LHC may be able to observe an
excess and discover vector-like quarks, or to instead strongly constrain VLQ
scenarios, by means of di-Higgs probes. We critically analyse 
existing bounds extracted from di-Higgs data, pointing out additional
information which can be obtained from the present searches. We moreover discuss
specific issues related to di-Higgs production in VLQ models, including the
relevance of boosted and non-boosted topologies, as well as the relative
strength of the QCD and electroweak (EW) production modes that could give rise
to a di-Higgs system from the decay of VLQs. We in particular investigate the
connection between the single and
the pair production of VLQ particles. We extend our preliminary work performed
in the context of the {\it 2015 Les Houches Workshop on TeV Collider
Physics}~\cite{Brooijmans:2016vro}, and additionally include next-to-leading
order (NLO) QCD effects in
a study of scenarios in which the sole non-vanishing EW VLQ coupling involves a
Higgs boson. Such effects are expected to be particularly important when
the VLQs couple to the first
generation quarks~\cite{Fuks:2016ftf,Brooijmans:2016vro}.

This article is organised as follows. Section~\ref{sec:model} contains a
description of the effective framework that we employ for our VLQ study, and
includes possibilities for linking it to ultraviolet-complete realisations.
This section also investigates di-Higgs production in VLQ models, and presents a
comprehensive analysis of the diverse production modes of two Higgs bosons in
such models. In Section~\ref{sec:searches}, we detail our simulation setup and
study the phenomenological consequences of our signal on existing LHC searches.
In Section~\ref{sec:signalcat}, we compare and contrast the VLQ-induced QCD and
EW production modes of a VLQ pair, and propose some variables that could be
useful to distinguish them. We give our conclusions in Section~\ref{sec:con}.

\section{Theoretical framework} \label{sec:model}
\subsection{Model-independent implementation}
\label{sec:theo}

We consider a simplified model where the SM is supplemented by new VLQs that only
couple to the Higgs boson and the SM quarks, on top of the usual QCD gauge
interactions. A model-independent parameterisation able to describe the main
features of a generic VLQ was proposed in Ref.~\cite{Buchkremer:2013bha}. Its
main advantage consisted in the free coupling parameters that were describing
both the VLQ branching ratios and single production cross sections. This model
has been recently extended so that QCD NLO effects can be included at
the VLQ production level~\cite{Fuks:2016ftf}.

In the following, we use the above-mentioned NLO model after having turned off
all unnecessary VLQ couplings to the EW gauge bosons. We restrict ourselves to
the cases of a top-like heavy quark $T$ (with an electric charge of 2/3)
of mass $M_{\sss T}$ and a bottom-like heavy quark $B$ (with an electric charge
of $-1/3$) of mass $M_{\sss B}$, as any other
extra quark would have an electric charge preventing it from coupling
to the Higgs boson. The simplified Lagrangian that we consider is therefore
given by
\be\bsp
 {\cal L}_{\rm VLQ} = &\ 
   i \bar B \slashed{D} B - M_{\sss B} \bar B B
 + i \bar T \slashed{D} T - M_{\sss T} \bar T T \\
 &\quad - h\ \bigg[
    \bar B  \Big(\hat\kappa_{\sss L}^{\sss B} P_L + \hat\kappa_{\sss R}^{\sss B}
       P_R\Big) q_d
  + \bar T \Big(\hat\kappa_{\sss L}^{\sss T} P_L + \hat\kappa_{\sss R}^{\sss T}
       P_R\Big) q_u
   + {\rm h.c.} \bigg]\,,
\esp\label{eq:LVLQ}\ee
where the covariant derivatives only include QCD interactions, and $P_L$ and
$P_R$ stand for the left-handed and right-handed chirality projectors. Although
weak gauge couplings could have been also introduced, their contribution is
always subleading, on top of being model-dependent, so that they have been
ignored~\cite{Buchkremer:2013bha}. We have however retained the couplings of
a single VLQ to the Higgs boson and a SM quark (SM flavour indices being
understood) on the second line of the above equation. While the NLO
implementation directly uses the $\hat\kappa$ parameters appearing in
Eq.~\eqref{eq:LVLQ} as free parameters, our phenomenological analysis relies on
the conventions of Ref.~\cite{Buchkremer:2013bha} where the couplings are fixed
as
\be \label{eq:kappa}
  \hat{\kappa}_{\sss L/R}^{\sss T/B} = \sqrt{2} \frac{M_{\sss T/B}}{v_{\rm SM}}
     \kappa_{T/B} \sqrt{\frac{\zeta_{L/R}^{i}}{2 \Gamma_h^0}}\,, 
  \qquad 2 \Gamma_h^0 = \left( 1 - \frac{M_h^2}{M_{T/B}^2} \right)^2\sim 1\,,
\ee
$v_{\rm SM}$ being the SM Higgs vacuum expectation value and $\kappa_{T/B}$ a
generic coupling strength. The rationale behind
this normalisation choice is such that the parameter $\zeta^i$ is equal to the
branching ratio associated with a $T/B$ decay into a Higgs boson and the
$i^{th}$ generation SM quark. The kinematic factor $\Gamma_h^0$ only plays a
role when a coupling to the $Z$ and/or $W$ boson is present, and it can
be approximated to $\Gamma_h^0 \sim 1/2$ for the VLQ mass range of interest. The
reason for the presence of a mass scaling in Eq.~\eqref{eq:kappa} arises from
the fact that $\hat{\kappa} \; v_{\rm SM}/M_Q$ is proportional to the mixing angle
between the VLQ and the SM quarks, which also affects the couplings of the SM
quarks.

It can be proven, in general, that the mixing angles are chiral. In other words,
the VLQ coupling to the Higgs boson and a specific chirality of SM fermion
is suppressed with respect to the coupling to the other chirality by the mass of
the SM fermion~\cite{Buchkremer:2013bha}. In any phenomenological analysis, one
has thus the freedom to choose either left or right-handed VLQ couplings.
The mixing between the extra quark and the SM quarks is however a constrained
quantity. It is bounded by experimental data that includes precision
measurements in the EW sector~\cite{Cacciapaglia:2011fx,Ellis:2014dza,Cacciapaglia:2015ixa},
flavour observables~\cite{Cacciapaglia:2011fx,Ishiwata:2015cga} and the
currently allowed room for deviations in the Higgs sector~\cite{Dawson:2012di,Bizot:2015zaa}.
These constraints on the mixing angles are
approximately independent on the mass of the VLQ because they come from lower
energy measurements. As an example, it was found that for a VLQ mixing to the
first generation only, as in the VLQ realisations considered in the rest of this
paper, the bounds mainly stem from deviations of the $Z$-boson
coupling to up quarks as measured in atomic parity violation experiments, and
leads to
$\kappa_T \lesssim 0.07$~\cite{Buchkremer:2013bha}. Even though the mixing angle
is in principle inversely proportional to the VLQ mass, we decided to fix its
value independently on the VLQ mass. With this choice, it turns out to be easier
to compare our findings with low energy bounds. The precise value of the
coupling $\kappa_{T/B}$, however, crucially depends on the specific model.

As illustrative examples, we briefly describe in the rest of this section two
scenarios where our model configuration featuring exclusive VLQ couplings to
the Higgs boson and a light SM quark arises naturally. One of these examples
concerns a weakly coupled theory and the other one a strongly coupled theory.

\subsection{Concrete examples}

The simplest example of a VLQ model featuring exclusive couplings to the Higgs
boson
and the first generation of SM quarks consists of an extension of the SM where
two weak doublets of extra quarks are added to the SM. The first one, denoted by
$Q_1 = (U_1, D_1)$, has a SM hypercharge of $Y_{\rm SM} = 1/6$ and the other
one, denoted by $Q_2 = (X_2, U_2)$, has an hypercharge
$Y = Y_{\rm SM} + 1 = 7/6$. Both doublets can couple to the SM fermions via
Yukawa interactions with the
Brout-Englert-Higgs field, so that the mass Lagrangian and the (physical) Higgs
boson interaction terms read
\be
  - \mathcal{L} \supset M_1\ \bar{U}_{1L} U_{1R} + M_2\ \bar{U}_{2L} U_{2R}
    + \Big( 1+\frac{h}{v_{\rm SM}} \Big)
          \Big(y_1 \bar{U}_{1L} + y_2 \bar{U}_{2L}\Big)
      \tilde{u}_R\,,
\ee
where we have neglected the Yukawa interactions of the first generation SM quark
due to their smallness. In our notations, $M_1$ and $M_2$ denote the VLQ mass
parameters, $y_1$ and $y_2$ the
strengths of their interaction with the Higgs boson multiplied by the SM vacuum
expectation value $v_{\rm SM}$ and $\tilde u_R$ is the SM right-handed up-quark
field.

The interest behind this model is that for degenerate VLQ masses, {\it i.e.} for
the symmetric setup in which $M_1 = M_2 = M$ and $y_1 = y_2 = y$, the new
physics impact on the
$Z$-boson couplings to the SM quarks nicely cancels. Such cancellations can also
be achieved in the more general case in which the mass parameters are different,
but where the strength of the $Z\bar{u}u$ coupling and the electroweak
precision observables agree with data at the price of introducing VLQ couplings
both to the $Z$ and Higgs bosons. The considered symmetric scenario has been
first proposed in the context of
Higgs-boson production~\cite{Atre:2013ap}, and then generalised~\cite{Atre:2008iu,Atre:2011ae}. As a detailed study of the constraints on the free
parameters can be found in Ref.~\cite{Cacciapaglia:2015ixa}, we restrict ourself
to a summary of the main features of the model. 

The quark mass eigenbasis $(T, T', u_R)$ is obtained by rotating the gauge
eigenbasis $(U_1, U_2, \tilde u_R)$ as follows,
\be\bsp
  \tilde{u}_R = \sin \varphi_R\, T_R + \cos \varphi_R\, u_R\,, \qquad
  U_{1R}+U_{2R} = \sqrt{2} ( \cos \varphi_R\, T_R - \sin \varphi_R\, u_R) \,, \\
  U_{1R} - U_{2R} = \sqrt{2} T'_R\,, \qquad
  U_{1L} + U_{2L} = \sqrt{2} T_L\,, \qquad
  U_{1L} - U_{2L} = \sqrt{2} T'_L\,,
\esp\ee
where $u_R$ is the SM (massless) quark field, $T$/$T'$ are the VLQ mass
eigenstates and $\varphi_R$ is the mixing angle of the heavy quark with the SM
quark. The physical masses are given by
\be
  M_T = \frac{M}{\cos\varphi_R} \qquad\text{and}\qquad
  M_{T'} = M = M_T \cos\varphi_R\ , \qquad \mbox{with}\qquad
  \tan \varphi_R = \frac{\sqrt{2} y}{M}\,. 
\label{eq:Mm1}
\ee
After these redefinitions, it is found that the heavier state $T$ couples to the
SM up-type quark exclusively via the Higgs boson, with an interaction strength
that reads
\be
  \hat{\kappa}^T_R = \frac{M_T}{v} \cos \varphi_R \sin \varphi_R \ .
\ee
We consequently derive from Eq.~\eqref{eq:kappa},
\be
 \kappa_T \sim \frac{\cos \varphi_R \sin \varphi_R}{\sqrt{2}} \,,
\label{eq:kaptr} \ee
in the notations of the simplified model that we have introduced in
Section~\ref{sec:theo}. Moreover, the model contains three lighter and
degenerate states,
namely the $T'$ up-type quark, a bottom-like heavy quark $B \equiv D_1$ and an
$X \equiv X_2$ state with an electric charge of 5/3. These three quarks couple
to the SM quarks and the massive weak gauge bosons with a strength proportional
to the sine of the mixing angle $\varphi_R$,
\be\bsp
  T'\scalebox{0.85}[1.0]{-}Z\scalebox{0.85}[1.0]{-}u  &\
    \Rightarrow \tilde{\kappa}^{T'}_R = - \sin \varphi_R\,, \\
  B\scalebox{0.85}[1.0]{-}W^-\scalebox{0.85}[1.0]{-}u &\
    \Rightarrow \kappa^B_R = - \frac{\sin \varphi_R}{\sqrt{2}}\,,\\
  X\scalebox{0.85}[1.0]{-}W^+\scalebox{0.85}[1.0]{-}u &\
    \Rightarrow \kappa^X_R = - \frac{\sin \varphi_R}{\sqrt{2}}\,, 
\esp\label{eq:kapbr}\ee
in the notations of Ref.~\cite{Buchkremer:2013bha}.

In order to assess the phenomenological viability of such a realisation,
constraints on the masses and couplings of the light new resonances can be
estimated from a CMS analysis that performed a search for single- and
pair-produced light-flavour quark partners~\cite{CMS:2016pul}. Considering the
production of a pair of $X$ and $B$ quarks via strong interactions followed by
an exclusive VLQ decay into a $Wq$ system, the resulting $WW$ plus jets
signature can be used to extract a bound on the masses of these two $B$ and $X$
states independently of the mixing angle $\varphi_R$.

We derive bounds by comparing theoretical predictions for the total production
rate $\sigma_{pp\rightarrow BB, XX}$ to the CMS exclusion shown
in Ref.~\cite{CMS:2016pul}, we get an $M_B = M_X > 945$~GeV constraint.
For masses above this limit, the most stringent and sole constraint arises from
searches for singly-produced $B$ quarks subsequently decaying into a $Wq$ system
that is produced by $u\bar{u}$ annihilation. The reason is twofold. First, the
corresponding CMS
search is charge-sensitive and specifically targets the presence of $W^-$ bosons
in the final state. Next, the other relevant single VLQ production option giving
rise to $W^-$ bosons involves an $\bar{X}$ resonance. The latter can only be
produced via $\bar{u} \bar{u}$ annihilation that is suppressed by the parton
densities, which renders the search ineffective. This motivates the development
of a future search for $W^+q$ VLQ decays, which could benefit from a
parton-density-enhanced $u u$ initial state.

\begin{figure}
\centering
\includegraphics[width=0.49\textwidth]{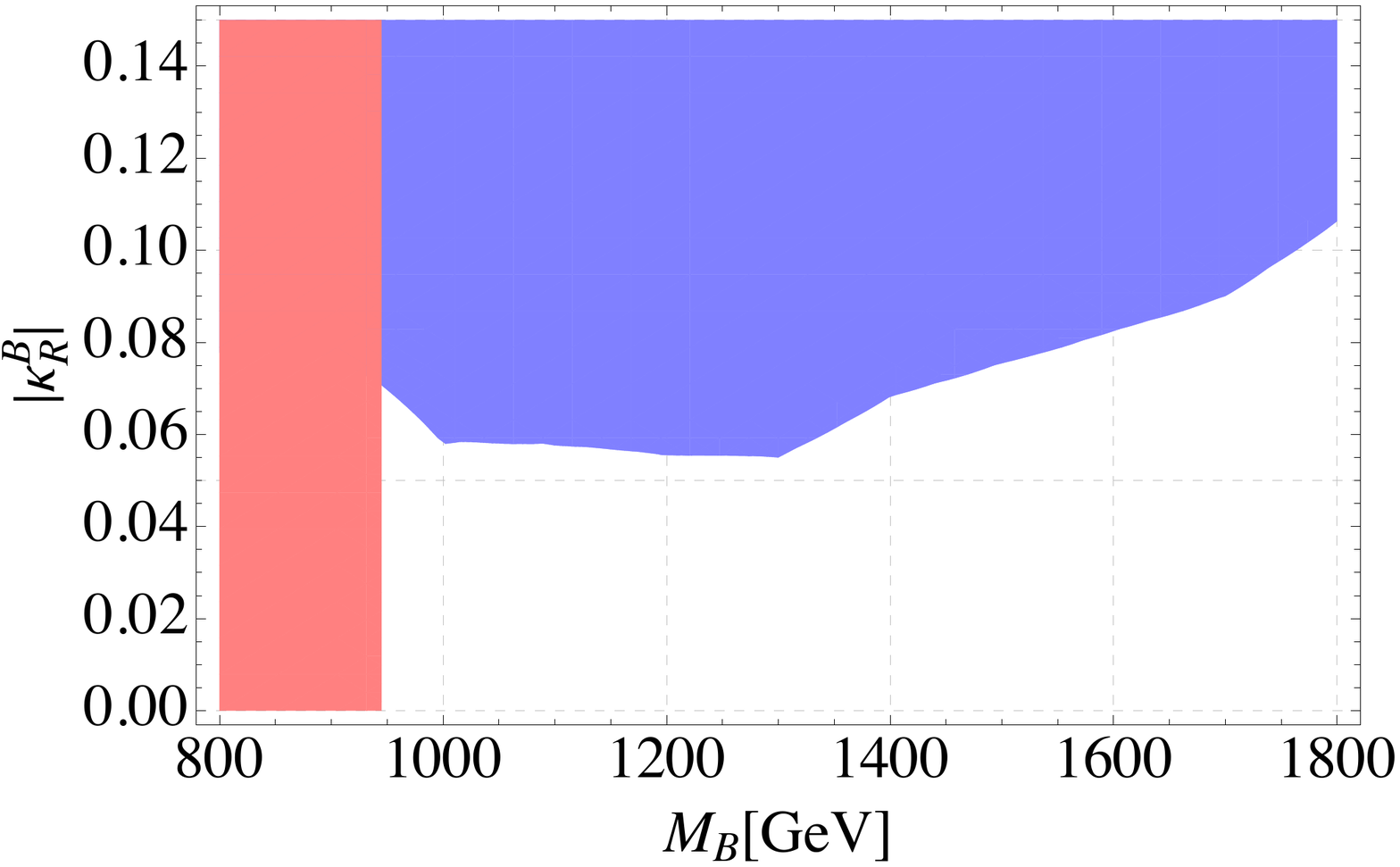}
\includegraphics[width=0.49\textwidth]{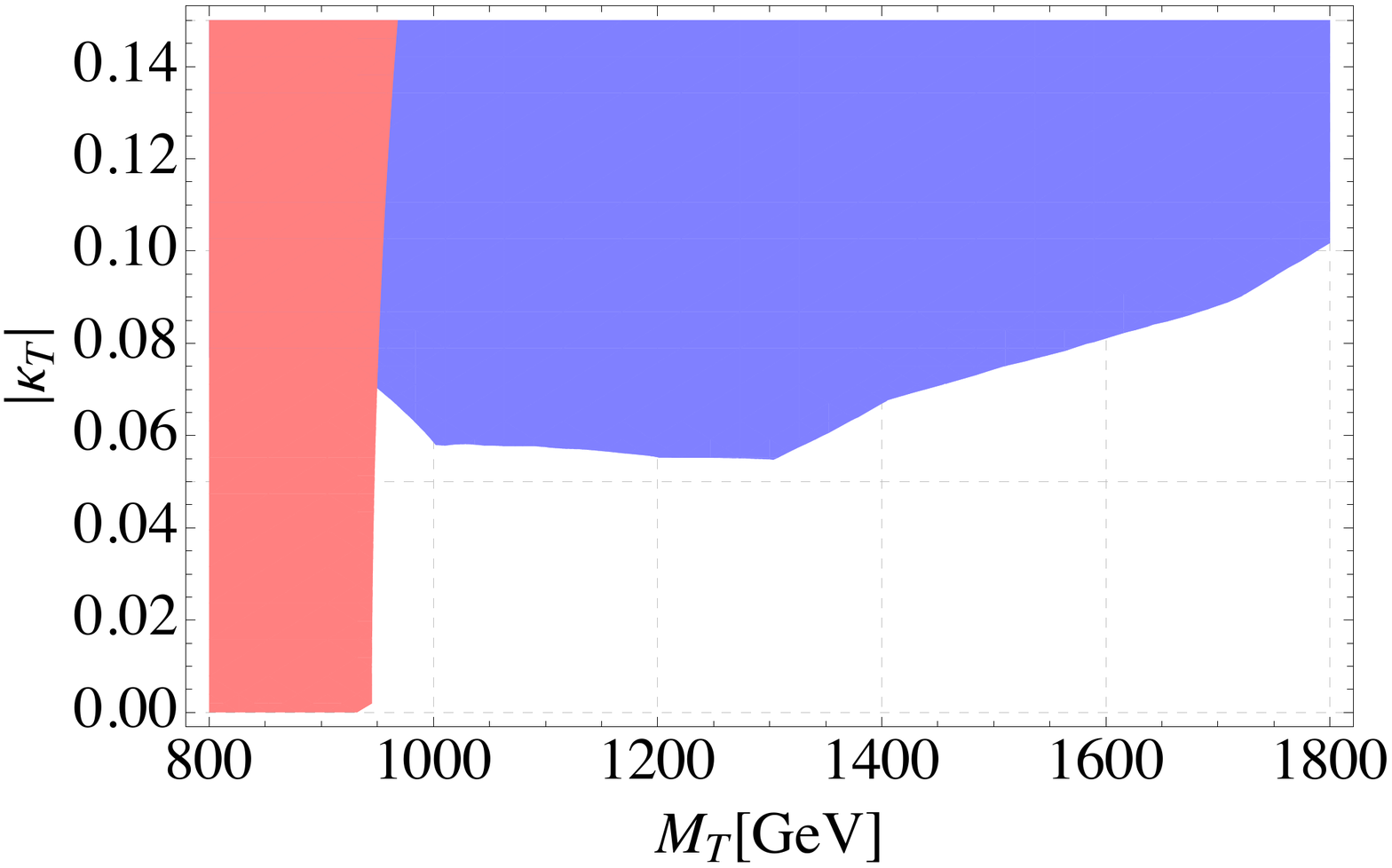}
\caption{\footnotesize Existing constraints on the parameters of the simplest
  VLQ realisation featuring extra quarks coupling to the first generation of SM
  quark and the Higgs boson, as
  extracted from the CMS search of Ref.~\cite{CMS:2016pul}. The bounds are
  equivalently shown in the $(\kappa^B_R, M_{\sss B})$ plane (left) and
  $(\kappa_T, M_{\sss T})$ plane (right). The red area is excluded by the search
  for a strongly produced pair of VLQs, while the blue area shows the additional
  exclusion arising from the search for a singly-produced $B$-quark.
 \label{fig:m1}
}
\end{figure}

Making use of the results of Ref.~\cite{CMS:2016pul} and Eq.~\eqref{eq:kapbr},
we derive a bound on the $\kappa^B_R$ coupling that depends on the $B$ quark
mass (left panel of Figure~\ref{fig:m1}). In the right panel of
Figure~\ref{fig:m1}, we re-express the results, via the mixing angle $\varphi_R$,
in terms of the $T$ coupling to the Higgs boson and the $T$-quark mass. The red
area indicates the bounds derived from VLQ pair production, and the blue region
reflects the single-VLQ production constraints. Further constraints could
however arise from EW precision tests~\cite{Cacciapaglia:2015ixa}.

The second considered example of models originates from Composite Higgs setups
with partial compositeness and an extended custodial symmetry~\cite{%
Agashe:2006at}. Independently on the symmetry breaking pattern, the minimal
top-partner content consists of a bi-doublet and a singlet field of extra quarks
charged under the custodial $SO(4)$ symmetry. Both the bi-doublet and the
singlet fields mix linearly with the SM elementary fields via two pre-Yukawa
couplings that we denote $y_L$ (for the bi-doublet) and $y_R$ (for the singlet),
while no SM Yukawa interaction is present. We choose a mixing with the first
generation of SM quarks. As the small mass of the up quark is proportional to
the product of the two $y_L$ and $y_R$ couplings, we assume that $y_L \ll y_R$
and hence decouple the bi-doublet from the rest of the
spectrum~\cite{Delaunay:2013pwa}.

In this limit, the mass and Yukawa interaction Lagrangian for the composite
singlet field $\tilde{U}$ reads
\be
  - \mathcal{L} \supset \bar{\tilde{U}}_L 
    \Big(M_1\ \tilde{U}_R + y_R f \cos \epsilon\ \tilde{u}_R\Big)\,,
\ee
where $\tilde u_R$ stands for the right-handed SM up-quark field and
$\sin \epsilon = v_{\rm SM}/f$ includes the effect of the EW symmetry breaking,
$f$ being the composite scale. A redefinition of the right-handed fields allows
for the rewriting of the Lagrangian in the mass eigenbasis, the heavy quark mass
$M_{\sss T}$ and mixing angle $\varphi_1$ being related to the Lagrangian
parameters as
\be
  M_T = \sqrt{M_1^2 + y_R^2 f^2 \cos^2 \epsilon} = \frac{M_1}{\cos \varphi_1}
  \quad\mbox{with}\quad \tan \varphi_1 = \frac{y_R f \cos \epsilon}{M_1}\,.
\ee
As this redefinition only involves right-handed fields that are not sensitive to
the weak interactions, no off-diagonal gauge interactions are generated.
However, an off-diagonal coupling to the Higgs boson arises from the
fact that the elementary right-handed up-quark couples to the composite object
non-linearly, {\it i.e.} via a cosine function. The corresponding interaction
term is in this case given by
\be
  \mathcal{L} \supset y_R \sin \epsilon \; h\ \bar{\tilde{U}}_L \tilde{u}_R
   = \frac{M_{\sss T}}{v_{\rm SM}} \frac{\sin^2 \epsilon}{\cos \epsilon}
     \frac{\sin 2 \varphi_1}{2}\; h\ \bar{T}_L u_R + \dots + {\rm h.c.}
\ee

Typically, the value of $\sin \epsilon$ is bounded to be small by EW precision
tests, so that we consider a generic bound of
$\sin^2 \epsilon \lesssim 0.1$~\cite{Cacciapaglia:2015ixa}. In the notations of
Eq.~\eqref{eq:kappa} and Eq.~\eqref{eq:kaptr}, the coupling strength of the VLQ
to the Higgs boson has therefore a natural upper bound of
\be
  \hat{\kappa}^T_R \lesssim \frac{M_{\sss T}}{v_{\rm SM}} \cdot 0.05
    \quad \Rightarrow \quad \kappa_T \lesssim 0.035\,.
\ee

In the limit where the singlet field decouples from the spectrum, one can
design an alternative composite realisation with the mixing of a bi-doublet to
the SM elementary quark field. This can be mapped to the example introduced at
the beginning of this section, with the equalities $M_1 = M_2$ and $y_1 = y_2$
being naturally enforced as the two doublets of the first example are now the
components of a single multiplet of the custodial $SO(4)$
symmetry~\cite{Delaunay:2013pwa}.

\subsection{Di-Higgs production at the LHC} \label{sec:signal}
\begin{figure}
  \centering
  \includegraphics[width=0.9\textwidth]{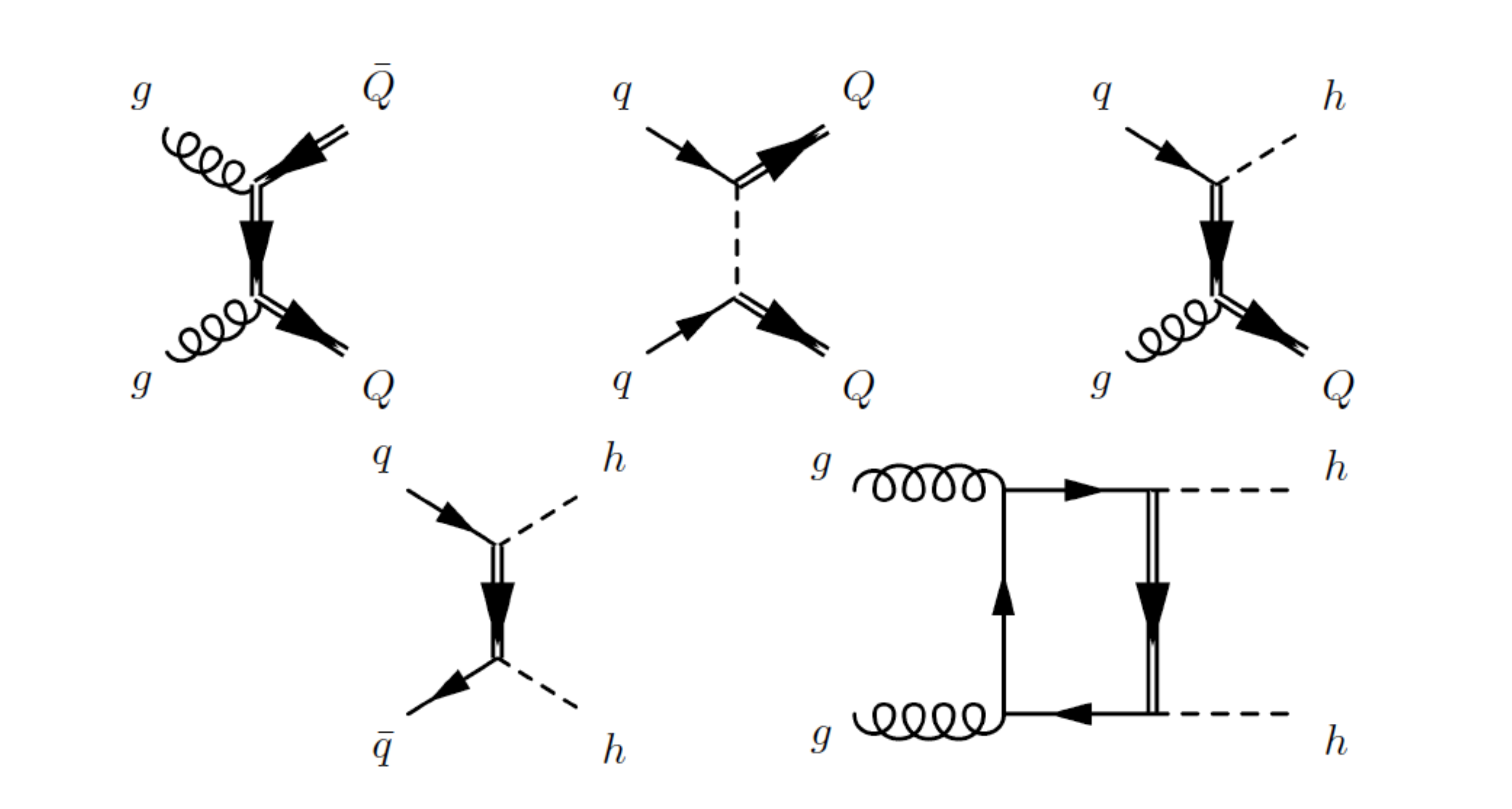}
  \caption{\footnotesize{Sample LO Feynman diagrams for each of the five
  production channels described in the text and giving rise a VLQ-induced
  di-Higgs final states, with the decay $Q \to h q$ not explicitly indicated.
  \label{fig:diagrams}}}
\end{figure}

Non-resonant di-Higgs production can result both from processes involving
internal VLQ propagators, or from the production of a single VLQ (in association
with a Higgs boson) or a pair of
VLQs that then decay into a Higgs boson and a jet. We describe in this
subsection all these relevant production modes in the context of VLQ setups
featuring a VLQ coupling to the Higgs boson with an up-type quark exclusively.
Di-Higgs states can be produced either alone or in association with jets, as
illustrated on Figure~\ref{fig:diagrams} where we show representative
leading-order (LO) Feynman diagrams for the five different production
mechanisms. In those diagrams, all external VLQs are understood to decay into a
Higgs boson and a jet.

Di-Higgs production can be induced by the production of a pair of
VLQs through strong interactions, as illutrated by the first diagram of
Figure~\ref{fig:diagrams}. This mechanism, being only sensitive to gauge
interactions, is independent of the value of the $\kappa_{T/B}$ parameters.
In this case, the di-Higgs final state arises from two $Q \to q h$ decays and is
thus produced in association with two additional jets, already at the LO
accuracy. As mentioned in Section~\ref{sec:theo}, additional contributions could
stem from EW gauge interactions, via, {\it e.g.}, $s$-channel $W$/$Z$ and photon
exchanges. These model-dependent contributions are however typically small, and
thus omitted.

Alternatively, the pair of VLQs could be produced electroweakly
via diagrams that depend on $\kappa_{T/B}^2$, as illustrated by the second graph
on Figure~\ref{fig:diagrams}. Although such contributions are naively thought to
be much smaller than the QCD corresponding processes, VLQ carrying the same
electric charge could be produced via such a mechanism, which benefits from
an enhancement originating from the parton densities (PDF). We have furthermore
verified that for the parameter space regions under consideration, interferences
of EW and QCD contributions are always negligible.

A pair of Higgs boson can also be produced from the associated production of a
single VLQ and a Higgs boson, the second Higgs boson arising from the VLQ decay.
The di-Higgs system is thus produced together with an extra hard-scattering jet,
and the corresponding amplitude, corresponding to the third Feynman diagram in
Figure~\ref{fig:diagrams}, scales linearly with $\kappa_{T/B}$.

The di-Higgs system can also be produced directly, the VLQ only appearing as
internal propagators. This is illustrated by the fourth and fifth diagrams
of Figure~\ref{fig:diagrams}. The first of these diagrams is representative of
di-Higgs production via a $t$-channel VLQ exchange and is proportional, at
the level of the LO amplitude, to $\kappa_{T/B}^2$. In contrast, the second
contribution is similar to the loop-induced SM production mode (that is included
in the calculation), VLQs being additionally allowed to appear in the loop as
studied in Ref.~\cite{Dib:2005re,Dawson:2012mk,Dolan:2012ac,Grober:2016wmf}.

\begin{figure}
  \centering
  \includegraphics[width=0.48\textwidth]{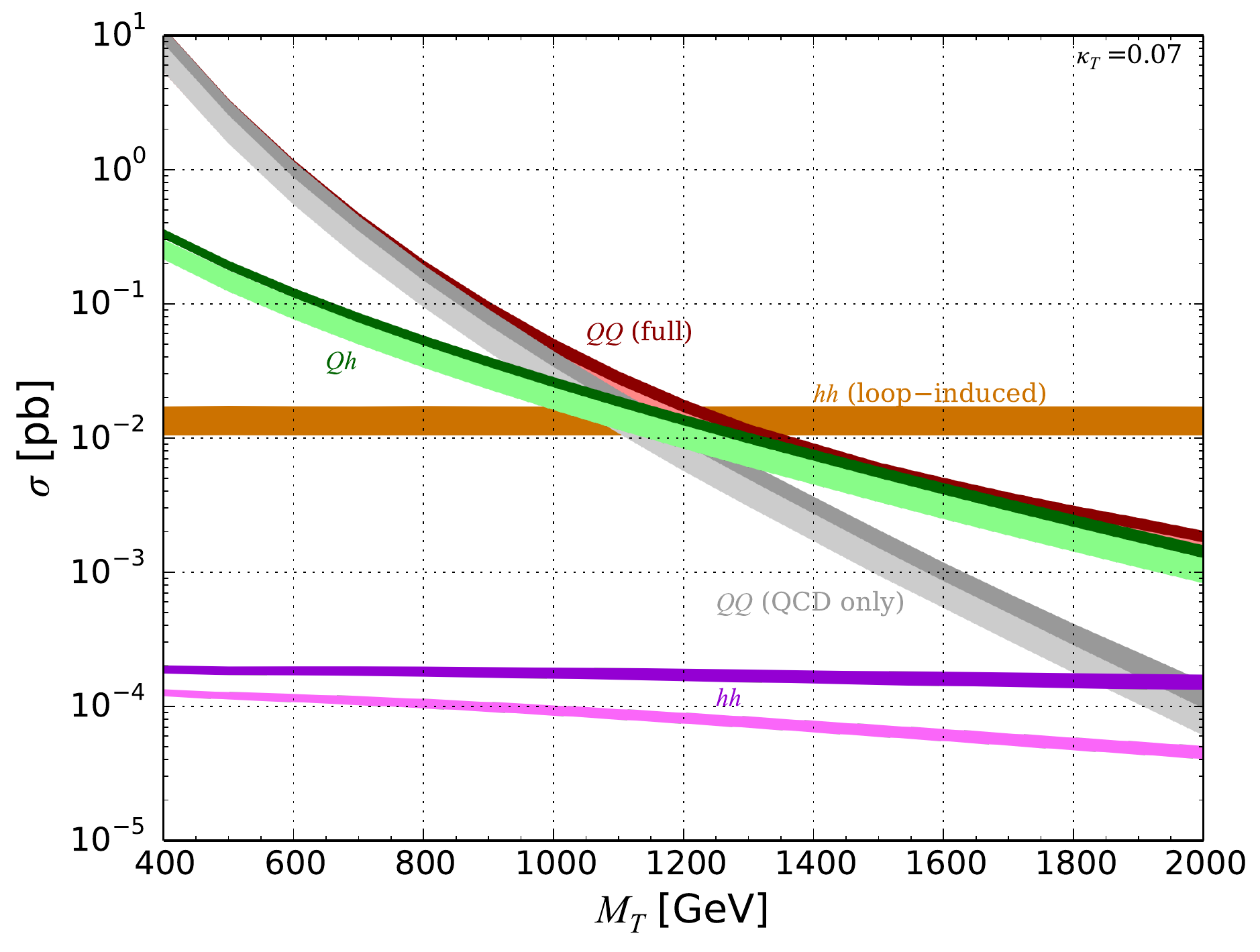}
  \includegraphics[width=0.48\textwidth]{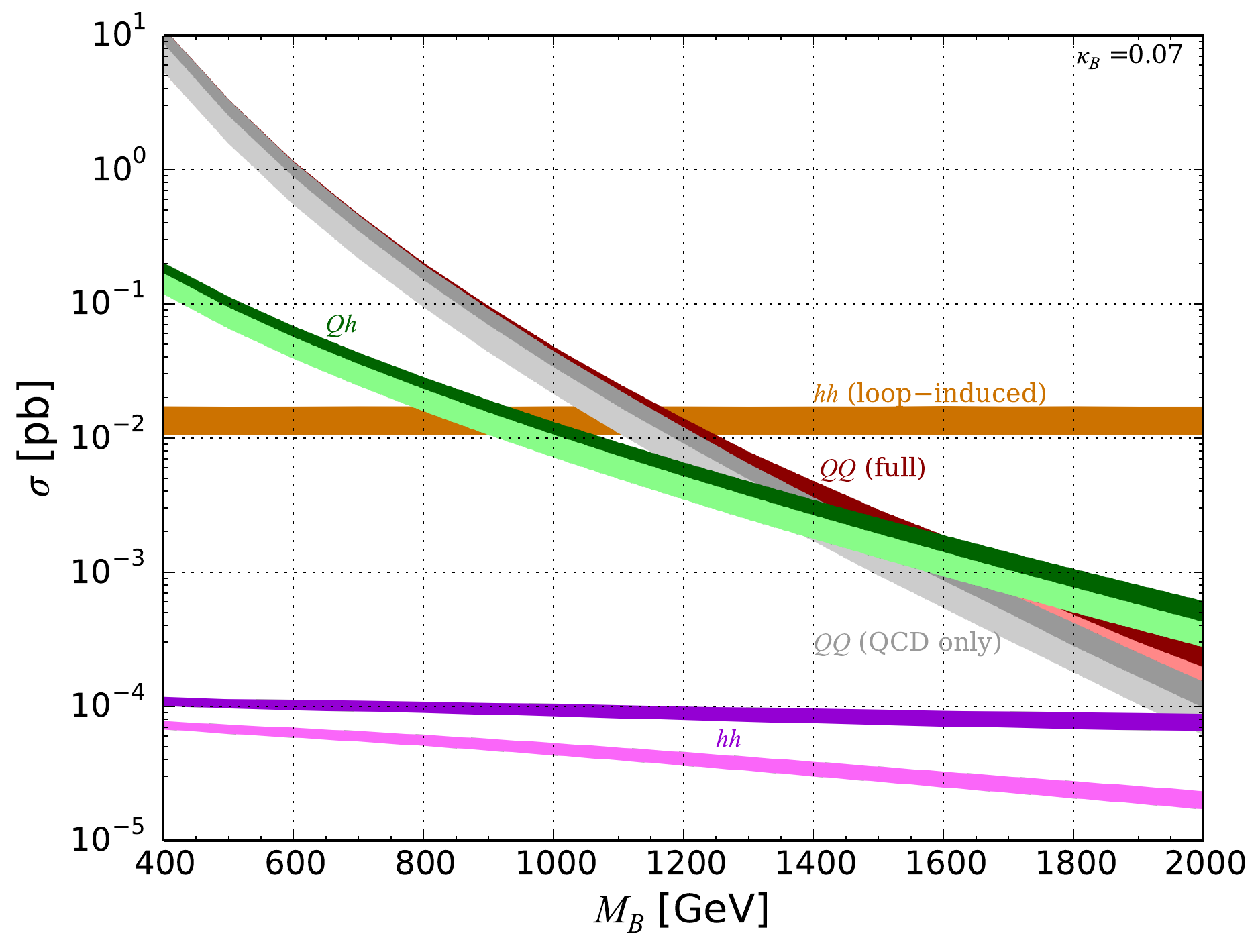}
  \caption{\footnotesize{Di-Higgs production cross sections at the LHC, for
  proton-proton collisions at a centre-of-mass energy of 13 TeV and for a
  benchmark scenario in which $\kappa_{T/B}=0.07$. The results are shown as a
  function of the VLQ mass and we consider up-type (left) and down-type
  (right) quark partners coupling to the first generation. A detailed
  explanation of the various curves is given in the text. LO results (light
  colours) and associated error bands can be compared to their NLO counterparts
  (darker colours).
  \label{fig:CX1D}}}
\end{figure}

In Figure~\ref{fig:CX1D}, we compare the total cross sections associated with
these different di-Higgs production modes for $\kappa_{T/B} = 0.07$. For each
production mechanism, we
include LO (lighter colours) and NLO (dark colours) predictions in QCD, with the
associated uncertainty band. For all subprocesses, NLO effects are found to
largely enhance the rates and reduce the errors, as emphasized in
Ref.~\cite{Fuks:2016ftf}. QCD-induced VLQ production contributions
(in grey) are independent on the quark flavour and drop quickly with the VLQ
mass. However, EW diagram contributions, that depend on the $\kappa_{T/B}$
coupling and are thus enhanced by the VLQ mass as shown by Eq.~\eqref{eq:kappa},
start to contribute for VLQ masses larger than about 1~TeV, as shown by the red
bands that include all ({\it i.e.} both the QCD and EW components) VLQ-induced
di-Higgs production processes $pp \to QQ,
Q\bar{Q}, \bar{Q} \bar{Q}$.  The effect is found more dramatic for up-type
quarks, which mainly results from the $p p \to Q Q$ process that is PDF-enhanced
as it could proceed via two up valence quarks.

The cross sections associated with $Qh$ production are depicted by green bands,
and benefits from a smaller phase space suppression for large VLQ masses than
for VLQ pair production. For the considered benchmark scenario in which
$\kappa_{T/B} = 0.07$, pair
and single VLQ production yield cross sections of similar values when QCD
contributions to VLQ pair-production become subleading, {\it i.e.} for
$M_T \gtrsim 1$~TeV. The phase-space suppression associated with the EW
contributions to VLQ pair-production is compensated by the coupling
enhancement.

\begin{figure}
  \centering
  \includegraphics[width=0.65\textwidth]{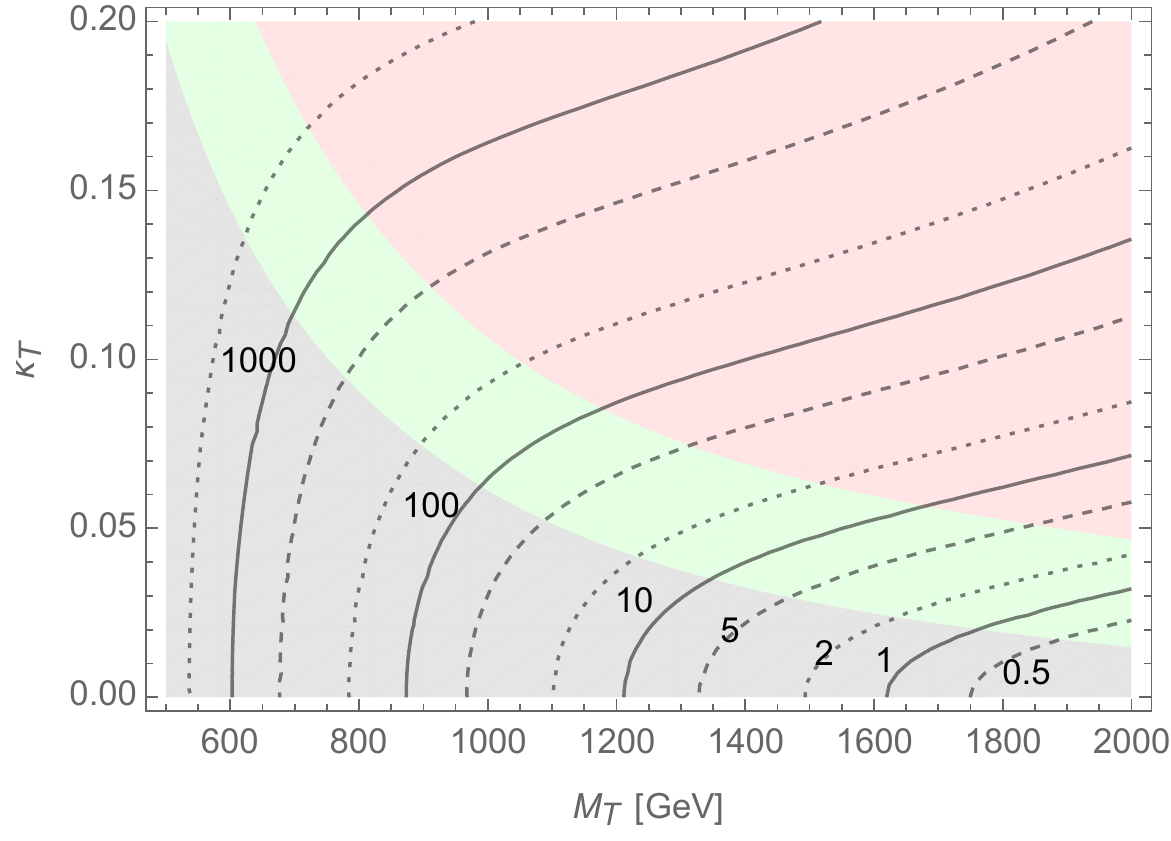}
  \caption{\footnotesize{Di-Higgs (total) production cross
  section at the LHC shown as contours in the $(M_Q, \kappa_Q)$ plane. The
  results, to be read in fb, contain both the single VLQ and VLQ-pair components
  and we represent as coloured regions the parameter space areas where a given
  production channel dominates (QCD-induced VLQ pair production in grey,
  EW-induced VLQ pair production in red and $Qh$-induced production in green).
  Direct (loop-induced and tree-level $t$-channel VLQ exchange) di-Higgs
  production has been neglected as the corresponding new physics contributions
  are always subleading for the parameter space regions under consideration.
\label{fig:CX2D}}}
\end{figure}

On the basis of the different cross section coupling scalings
detailed above, $Qh$-induced di-Higgs production only dominates over the $QQ$
mode for intermediate masses, EW pair production dominating instead
for very large masses.
This is further illustrated in Figure~\ref{fig:CX2D} where total cross section
contours, summed over all pair and single production modes, are shown in the
$(M_Q, \kappa_Q)$ plane. We distinguish the different regions of the parameter
space in terms of the dominant di-Higgs production mode. Cases dominated by the
QCD contributions to VLQ-pair production are shown in grey, by the EW
contributions to VLQ pair production in red, and by $Qh$ single production in
green. The latter is, as expected, found to only dominate in the intermediate
region where the VLQ mass is not too large (so that EW contributions are still
subleading) and not too small (to prevent the strong contributions from
dominating).

Loop-induced contributions to di-Higgs production by gluon fusion
(last diagram of Figure~\ref{fig:diagrams}), as well as direct di-Higgs
production via a $t$-channel VLQ exchange (fourth diagram of
Figure~\ref{fig:diagrams}) are always negligible in the parameter space regions
under consideration.
The $t$-channel VLQ exchange is shown as violet bands in Figure~\ref{fig:CX1D},
and the gluon fusion
loop-induced contributions are shown as a brown band. The former is subleading
and hardly contributes, even for large masses for which all other modes start to
be phase-space suppressed. This is still true even after accounting for the
gigantic NLO $K$-factors due to the quark-gluon-initiated contributions that
arise at NLO and that turn to dominate. The loop-induced results (brown) are
dominated by the SM top-loop diagram and are thus not depending on the VLQ setup
at all. These two
direct di-Higgs production channels however yield Higgs bosons possessing a low
transverse momentum $p_T$, so that the corresponding events are unlikely to populate
the signal regions of the corresponding analyses that target high $p_T$ Higgs
boson pair-production (see Section~\ref{sec:searches}). We leave for future work
the design of a specific study allowing to probe these production modes and
their specific topology.

\section{LHC searches at 13 TeV}\label{sec:searches}

Studying the pair-production of Higgs bosons at the LHC is an important
topic of research. There exist a large number of dedicated ATLAS and CMS
analyses at a centre-of-mass energy of $\sqrt{s} = $ 8 and 13~TeV, and they
differ by the final state they focus on. Various signatures have been
considered, which include final states with two \Hbb decays~\cite{%
ATLAS:2016ixk,Aaboud:2016xco,CMS:2016foy,CMS:2016tlj,%
CMS:2016pwo}, as well as those with one \Hbb and one \Htautau
decay~\cite{CMS:2016knm,CMS:2016ymn}, one \Hbb and one \Hgamgam decay~\cite{%
ATLAS:2016lala,CMS:2016vpz}, one \Hbb and one \HWW decay~\cite{CMS:2016cdj}, as
well as one
\HWW and one \Hgamgam decay~\cite{ATLAS:2016qmt}. Here we focus on the ATLAS and
CMS analyses using 13\,TeV data in the four-$b$-jet final state and estimate the
LHC sensitivity to the model proposed in Section~\ref{sec:model}.\footnote{For a recent study of the LHC high luminosity reach for SM di-Higgs searches in the four-$b$-jet final state we refer to ref. \cite{deLima:2014dta}.} This
final state has the advantage of arising from Higgs boson decays with the
largest branching fraction. In addition, $b$-jet tagging and, at high VLQ
masses, jet substructure techniques make this channel promising and worthy to
explore.

Both LHC collaborations have searched for new physics connected to di-Higgs
production in cases where the Higgs-boson pair is resonantly produced from the
decay of a heavier new particle, as well as when it is produced non-resonantly.
Searches are typically organised into two classes depending on the \pt of the
\Hbb system. The `resolved' final state analyses are optimised for
reconstructing low \pt Higgs boson pairs using four separate $b$-jets in the
detector, whereas `boosted' final state analyses target the high \pt \Hbb decays
where the showered and hadronised $b$ quarks are merged into one fat jet that is
clustered using a larger distance parameter than a more conventional jet arising
from a single hard parton.

In this section we use event selections inspired by
the corresponding ATLAS~\cite{Aaboud:2016xco} and CMS~\cite{%
CMS:2016tlj,CMS:2016pwo} analyses at $\sqrt{s} = 13$\,TeV to identify resolved
and boosted $hh\to4b$ configurations. As the experimental searches of
Refs.~\cite{Aaboud:2016xco,CMS:2016tlj,CMS:2016pwo} are mainly aiming at
the resonant production of a pair of Higgs bosons from a heavy new physics state
decay, both Higgs bosons are studied in the same way, {\it i.e.} either both
resolved or both boosted. The intermediate regime in which only one of the Higgs
bosons would be boosted is not considered, although it may be interesting when
the Higgs-boson system arises asymmetrically, as discussed in
Section~\ref{sec:signalcat}.

We estimate the possible
impact of the exotic production of a Higgs-boson pair through the decay of
intermediate vector-like quarks. The simulations employed in this section are
for an up-type vector-like quark generically denoted by $Q$. However, the VLQ
flavour does not affect the signal topologies for the final states under
consideration~\cite{Brooijmans:2016vro}.

\subsection{Simulation setup}\label{ss:Simulation}

For the simulation of the VLQ signal, we allow the VLQ mass $M_Q$ to vary and be
equal to 500\,GeV, 650\,GeV, 800\,GeV, 1\,TeV and 2\,TeV.
Signal simulation is performed within the \amc\ framework~\cite{Alwall:2014hca} where the
entire event generation process is automated~\cite{Christensen:2009jx}. We make
use of UFO model files~\cite{Degrande:2011ua} extracted from the Lagrangian
presented in Ref.~\cite{Fuks:2016ftf} with the help of the \fr~\cite{%
Alloul:2013bka}, \nloct~\cite{Degrande:2014vpa} and \fa~\cite{Hahn:2000kx}
packages, as this Lagrangian embeds the model introduced in
Section~\ref{sec:model}. Hard scattering events are generated at the NLO
accuracy in QCD, the virtual one-loop contributions being evaluated with the
\ml\ module~\cite{Hirschi:2011pa} and then combined with the real contributions
by means of the FKS subtraction method~\cite{Frixione:1995ms} as implemented in
the \mfks\ package~\cite{Frederix:2009yq}. VLQ decays are then handled
automatically by means of the \msp~\cite{Artoisenet:2012st} and
\mw~\cite{Alwall:2014bza} programs, which allow for retaining both off-shell and
spin correlation effects. The fixed-order calculations, for which the NLO set of
NNPDF 3.0 parton densities~\cite{Ball:2014uwa} is used, are matched with the parton shower and hadronization infrastructure of the \py\ package~\cite{Sjostrand:2014zea},
and we simulate the response of an LHC-like detector by using the \de\ program~\cite{deFavereau:2013fsa} (using a CMS or an ATLAS description where relevant) that internally relies on the \fj\ package~\cite{Cacciari:2011ma} for jet reconstruction.

\subsection{Resolved analyses\label{ss:ResolvedAna}}

We consider both an ATLAS-like and a CMS-like resolved di-Higgs boson analysis where all Higgs decay products can be entirely reconstructed~\cite{Aaboud:2016xco,CMS:2016tlj}.  In the following, a {\it resolved jet} denotes a jet candidate reconstructed by means of the anti-$k_t$ jet
algorithm~\cite{Cacciari:2008gp} with a distance parameter $R$ set to 
\be
  R = 0.4 \qquad (\text{also known as an AK4-jet}).
  \label{eq:AK4}
\ee
We additionally impose that the jet transverse momentum $\pt^j$ and
pseudorapidity $\eta^j$ satisfy
\be
  p_T^j > 20~{\rm GeV} \qquad \text{and} \qquad |\eta^j|< 2.5 \ .
  \label{eq:AK4PtEta}
\ee
Jets are potentially tagged as $b$-jets with an efficiency extracted from the maps provided
in Ref.~\cite{Chatrchyan:2012jua} and we additionally constrain the transverse
momentum of all $b$-tagged jets to fulfil
\be
 p_T^b > 40~{\rm GeV}\; ~\mbox{(ATLAS)}
  \qquad\text{or}\qquad
 p_T^b > 30~{\rm GeV}\; ~\mbox{(CMS)}
\ee
in our ATLAS-like and CMS-like analysis, respectively.
In both our resolved analyses, we select events that contain at least four
resolved $b$-tagged jets,
\be
  N(b) \geq 4 \ .
\ee
The four leading $b$-jets are then combined into two pairs of jets for which
the angular distance in the transverse plane obeys 
\be
  \Delta R (b,b) < 1.5\ ,
  \label{eq:DRbbCMS}
\ee
each pair of $b$-jets being assumed to originate from the decay of a Higgs
boson.

In the CMS-like analysis, we follow the {\it medium-mass region} selection~\cite{CMS:2016tlj} that has been designed to optimally probe resonantly produced di-Higgs systems where the resonance mass lies between 400\,GeV and 1200 GeV. Denoting by $M_{h_1}$ and $M_{h_2}$ the invariant masses of the two reconstructed Higgs boson candidates, we select events for which these masses satisfy
\be
  \chi_{\rm CMS}^2 =  \bigg( \frac{M_{h1}-\bar M_h}{\sigma_h}\bigg)^2
    + \bigg( \frac{M_{h2}-\bar M_h}{\sigma_h} \bigg)^2 < 1 \ ,
\label{eq:chi2CMS}\ee
where $\bar M_h$ is the average mean of the $M_{h_1}$ and $M_{h_2}$
distributions and is equal to 115\,GeV. We moreover choose a width $\sigma_h=23$\,GeV,
as stemming from the CMS procedure for increasing the analysis sensitivity in
the medium-mass region.

In contrast, our ATLAS-like analysis includes first a selection on the
transverse momentum, pseudorapidity and invariant mass of the reconstructed
Higgs bosons. The transverse momentum of the leading reconstructed Higgs-boson
candidate $p_T(h_1)$ is constrained to fulfil
\be
  p_T(h_1) > \left\{ \begin{array} {l l}
     200~{\rm GeV} \qquad & \text{for}~m_{4j} < 600~{\rm GeV} \ ,\\
     0.65\ m_{4j} - 190~{\rm GeV} \qquad & \text{for}~m_{4j}
       \in [600, 910]~{\rm GeV} \ ,\\
     400~{\rm GeV} \qquad & \text{for}~m_{4j} > 910~{\rm GeV} \ ,
  \end{array}\right.
\ee
where $m_{4j}$ is the invariant mass of the system made of the two reconstructed
Higgs bosons (or equivalently of the four leading $b$-jets), while the
transverse momentum of the subleading Higgs-boson candidate $p_T(h_2)$ must obey
\be
  p_T(h_2) > \left\{ \begin{array} {l l}
     150~{\rm GeV} \qquad & \text{for}~m_{4j} < 520~{\rm GeV} \ ,\\
     0.23\ m_{4j} + 30~{\rm GeV} \qquad & \text{for}~m_{4j}
       \in [520, 990]~{\rm GeV} \ ,\\
     260~{\rm GeV} \qquad & \text{for}~m_{4j} > 990~{\rm GeV} \ .
  \end{array}\right.
\ee
Moreover, the two reconstructed Higgs bosons are required to be not too
separated in pseudorapidity,
\be
  |\Delta\eta(h_1, h_2)| < \left\{ \begin{array} {l l}
    1 \qquad & \text{for}~m_{4j} < 820~{\rm GeV} \ ,\\
    0.0016\ m_{4j} - 0.28  \qquad & \text{for}~m_{4j} > 820~{\rm GeV} \ .
  \end{array}\right.
\ee
A final selection is imposed on the two masses of the reconstructed Higgs
bosons $M_{h_1}$ and $M_{h_2}$,
\be
  \chi_{\rm ATLAS}^2 =  \bigg( \frac{M_{h_1}-\bar M'_h}{0.1\ M_{h_1}}\bigg)^2
    + \bigg( \frac{M_{h_2}-\bar M_h}{0.1\ M_{h_2}} \bigg)^2 < 2.56 \ ,
\label{eq:ATLASresolved}\ee
with $\bar M'_h = 124$\,GeV.

\subsection{Boosted analyses}\label{ss:BoostedAna}

We  describe in this section our CMS-like and ATLAS-like boosted analysis of a
potential new physics di-Higgs signal. We denote as a CMS-like `fat jet'
($J_h^C$ with $C$ standing for CMS) any jet candidate reconstructed by means of
the anti-$k_t$ algorithm with a distance parameter fixed to
\be
   R = 0.8 \qquad (\text{also known as an AK8-jet}).
\ee
The $J_h^C$ candidate is tagged as a Higgs boson jet if it further satisfies
\be
  \big|\eta^{J_{h}^C}\big| < 2.4 \ , \qquad
  \tau_{21}^{J_{h}^C} < 0.6
  \qquad\text{and}\qquad
  m^{J_{h}^C}_{\rm pruned} \subset [105,135]~{\rm GeV}\ ,
\ee
where $\tau_{21}^{J_{h}^C}$ is the ratio of the $N=2$ and $N=1$ $N$-subjetiness variables~\cite{Thaler:2010tr} and $m^j_{\rm pruned}$ is the pruned jet mass~\cite{Ellis:2009me}. The efficiency associated with the $b$-tagging of a Higgs fat jet is implemented by assuming that the efficiency of tagging a subjet as a $b$-jet is the same as for a resolved jet with a transverse-momentum equal to half the fat jet transverse momentum. We have moreover optimistically assumed that any $b$-tagged fat jet contains two $b$-tagged subjets and we
select events that feature at least two $b$-tagged Higgs fat jets $B_h$,
\be
  N(B_h) \geq 2 \ .
\ee
The two leading $p_T$ CMS-like fat-jet are recognised as the two reconstructed
Higgs bosons $h_1$ and $h_2$, and we further enforce
\be
 |\eta(h_{1,2})| < 2.4\ ,
 \qquad
 p_T(h_{1,2}) > 200~{\rm GeV} \qquad\text{and}\qquad
 |\Delta\eta(h_1, h_2)| < 1.3\ .
 \label{eq:CMSDeta}
\ee
The final discriminant variable in the CMS-like boosted analysis is the reduced
mass
\be
  m_{\rm red} = m_{4j} - \Big(M_{h_1}-M_h\Big) - \Big(M_{h_2}-M_h\Big)
\label{eq:mred}\ee
with $M_h = 125$\,GeV and $m_{4j}$ abusively denoting the invariant mass of the
di-Higgs system (for having consistent notations with the previous section). The
reduced mass $m_{\rm red}$ is further imposed to be greater than 1~TeV.

For the ATLAS-like analysis, the `fat jet' $J_{h}^A$ (with $A$ standing for
ATLAS) is defined by once again using the anti-$k_t$ jet algorithm but with this
time a distance parameter set to
\be
   R = 1.0\ .
\ee 
While CMS uses boosted jet algorithms based on particle-flow tracks~\cite{%
Khachatryan:2014vla,CMS-PAS-PFT-09-001}, the ATLAS collaboration reconstructs
its fat jets (also called large-$R$ jets~\cite{ATLAS-CONF-2014-018}) from the
information extracted from the topological clusters of the hadronic calorimeter.
In our ATLAS-like boosted analysis, large-$R$ jets are clustered within the
{\sc FastJet} version embedded into \de\,
and then trimmed~\cite{Krohn:2009th}, before we apply constraints on the
invariant mass ($m^{J_{h}^A}$), pseudorapidity and transverse momentum of the
two leading fat jets,
\be
  m^{J_{h}^A} >  50~{\rm GeV}\ ,\qquad
 |\eta(J_{h}^A)| < 2
 \qquad\text{and}\qquad p_T(J_{h}^A)| > 250~{\rm GeV}.
\ee 
The signal region is defined by imposing extra constraints on the two
leading fat jets that are identified as the two reconstructed Higgs bosons
$h_1$ and $h_2$,
\beq    
  p_T(h_1) > 350~{\rm GeV}\ , \quad
  p_T(h_2) > 250~{\rm GeV} \,
  \quad \text{and}\quad
  |\Delta\eta(h_1, h_2)| < 1.7 \ ,
 \eeq
and by enforcing the $\chi^2_{\rm ATLAS}$ variable defined in
Eq.~\eqref{eq:ATLASresolved} to fullfil
\beq 
 \chi^2_{\text{ ATLAS}} < 2.56\ .
\eeq
After these selection, we derive the invariant mass of the reconstructed
Higgs bosons pair $m_{2J}$ that is  used in the ATLAS analysis, to
characterise the signal.

\subsection{Efficiencies and expectations}

\begin{figure}
  \centering
  \includegraphics[width=0.48\textwidth,valign=t]{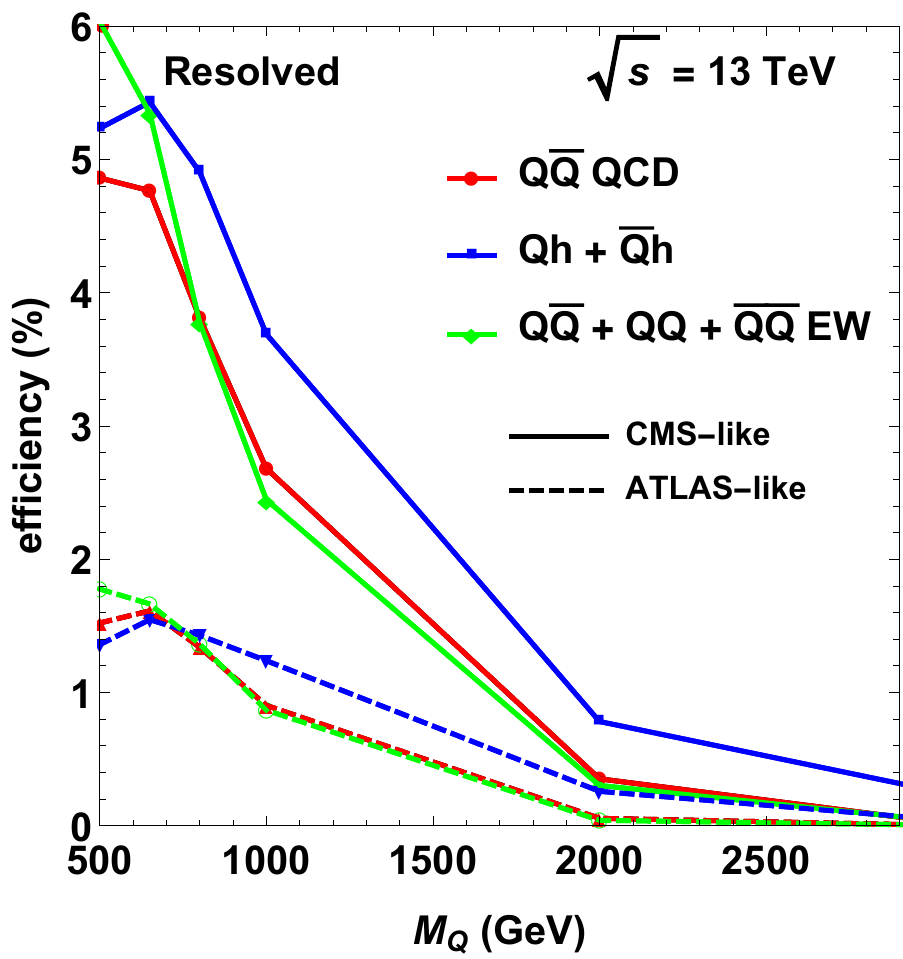}
  $~~$
  \includegraphics[width=0.48\textwidth,valign=t]{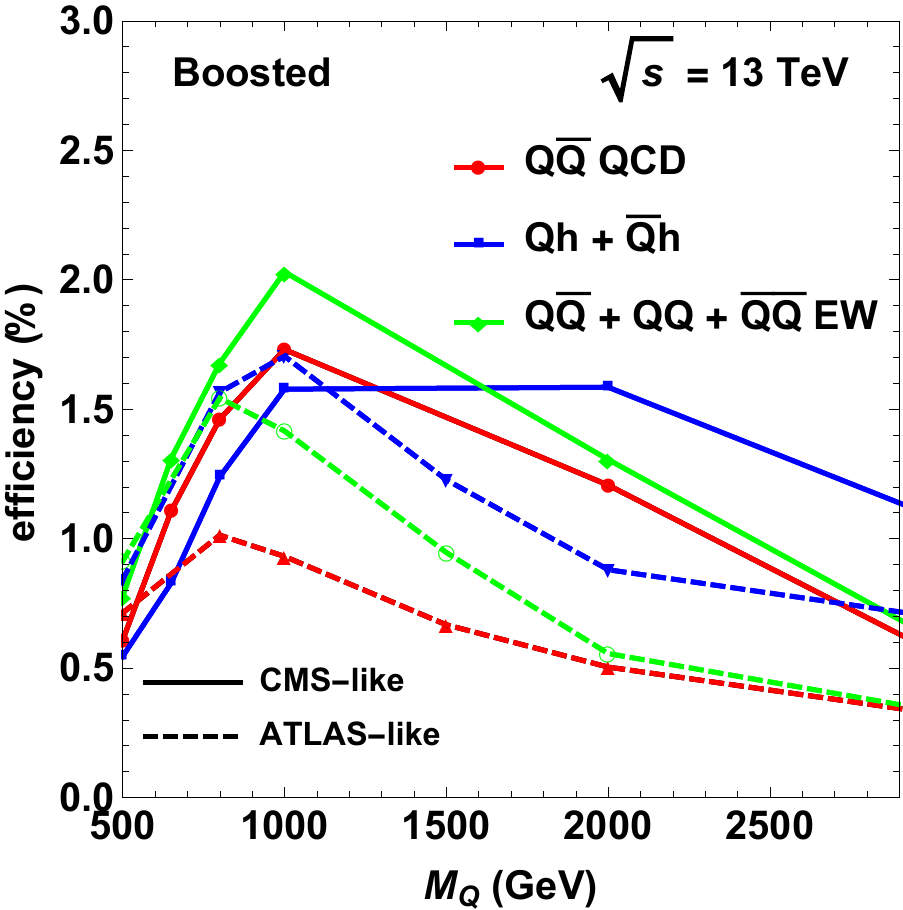}
  \caption{\footnotesize Selection efficiencies associated with the analyses
    introduced in Section~\ref{ss:ResolvedAna} and Section~\ref{ss:BoostedAna}
    for the resolved (left panel) and boosted (right panel) cases. The CMS-like
    and ATLAS-like strategies are respectively depicted by solid and dotted
    lines, and the results are based on the (NLO-QCD accurate) simulation of
    VLQ-induced di-Higgs production in proton-proton collisions at a
    centre-of-mass energy of $\sqrt{s}=13$\,TeV. We separately consider the QCD
    (red) and EW (green) production of a VLQ pair that then decays into a
    di-Higgs (plus two jets) system and the associated production of a single
    VLQ (that then decays into a Higgs boson plus jet system) with a
    Higgs boson (blue).
    \label{fig:RecastEfficiencies}
  }
\end{figure}

The analyses introduced in Section~\ref{ss:ResolvedAna} and
Section~\ref{ss:BoostedAna} are used to investigate the di-Higgs signal that
arises from the presence of VLQs in the context of the model described in
Section~\ref{sec:model}.
Figure~\ref{fig:RecastEfficiencies} shows the different signal selection
efficiencies that we present as a function of the VLQ mass. We consider both the
CMS-like (solid lines) and the ATLAS-like (dashed lines) analyses, and both the
boosted (right panel) and the resolved (left panel) selections. The different
sets of curves correspond to the QCD (red) and EW (green) production of a VLQ
pair (followed by two $Q\to h j$ decays), and to the associated production of a
vector-like quark (that then decays into a Higgs boson plus jet system) with a
Higgs boson (blue). In order to assess how well our
analyses match what could be expected from the corresponding experimental
analyses, we simulate a $pp\to G \to hh$ signal where $G$ is a Kaluza-Klein
graviton~\cite{Goldberger:1999wh,Oliveira:2014kla} and compare our findings to
the signal efficiencies presented in the original experimental publications. We
obtain an agreement at the sub-percent level.

In the context of the resolved analysis, we observe that the efficiencies
related to the QCD and electroweak VLQ pair production modes are almost
identical for a given quark partner mass as long as the VLQ is not too light, so
that the inclusion of the electroweak channels only modifies the total event
rate. As expected, the efficiency is maximal for light VLQ masses, reaching up
to 5\% for
the CMS-like analysis, and then decreases rapidly with increasing VLQ masses.
The efficiencies associated with the ATLAS-like analysis are also found much
lower, which is explained by the more severe selection strategy. For
$M_Q > 800$~GeV, the single VLQ production channel leads to efficiencies that
are higher
than for pair production. The inclusion of this channel is therefore useful to
assess the LHC sensitivity to VLQ models by means of di-Higgs probes more
accurately. A stronger reach can hence be expected, as already suggested by the
results shown in Figure~\ref{fig:CX2D}.

\begin{figure}
  \centering
  \includegraphics[width=0.49\textwidth]{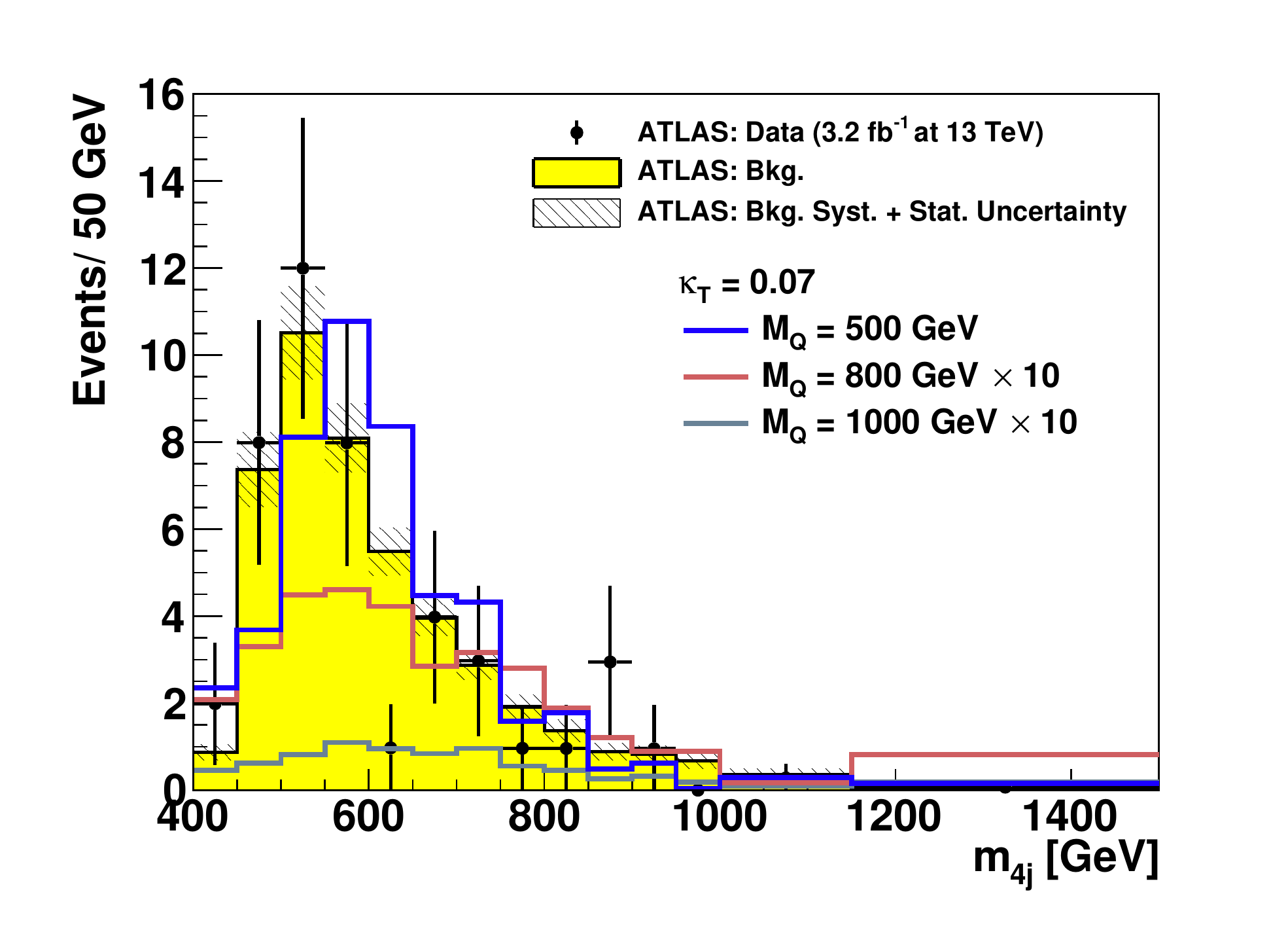}
  \includegraphics[width=0.49\textwidth]{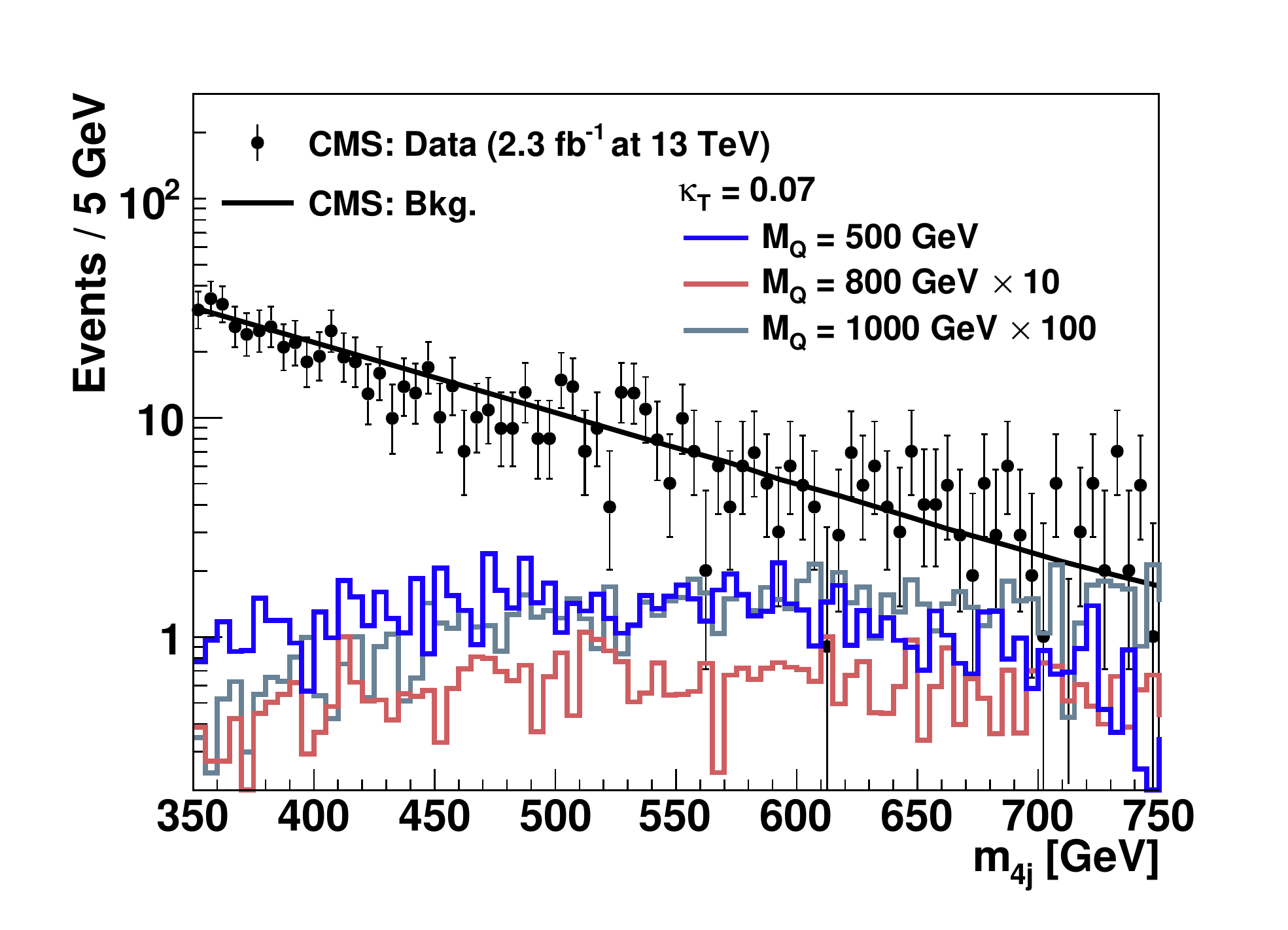}
  \caption{\footnotesize Distribution in the invariant mass of the reconstructed
  di-Higgs system both for the resolved ATLAS-like analysis (left) and CMS-like
  analysis (right). We select three benchmark setups with VLQ masses of 500\,GeV
  (blue), 800\,GeV (red) and 1\,TeV (grey) and with $\kappa_{\rm T} = 0.07$, and
  we overlay our predictions with the
  ATLAS~\cite{Aaboud:2016xco} and CMS~\cite{CMS:2016tlj} data and background.}
  \label{fig:ATLAS_CMS_resolvedRecast}
\end{figure}

The efficiencies originating from the boosted analyses are in general smaller
than for the resolved case, with a maximal values reached for $M_{Q} \sim 1$~TeV
and a range spanning 1--2\%. Genuine differences also appear when the di-Higgs
system originates from QCD or EW VLQ pair production, as these two processes
lead to a very different jet activity to which the boosted jet tagger is very
sensitive to. We moreover observe smaller efficiencies for the
ATLAS-like analysis due to the more severe selection strategy. These lower
efficiencies are however
not problematic as in the boosted regime, the background is also expected to be
much more reduced than in a resolved context.

\begin{figure}
  \centering
  \includegraphics[width=0.48\textwidth]{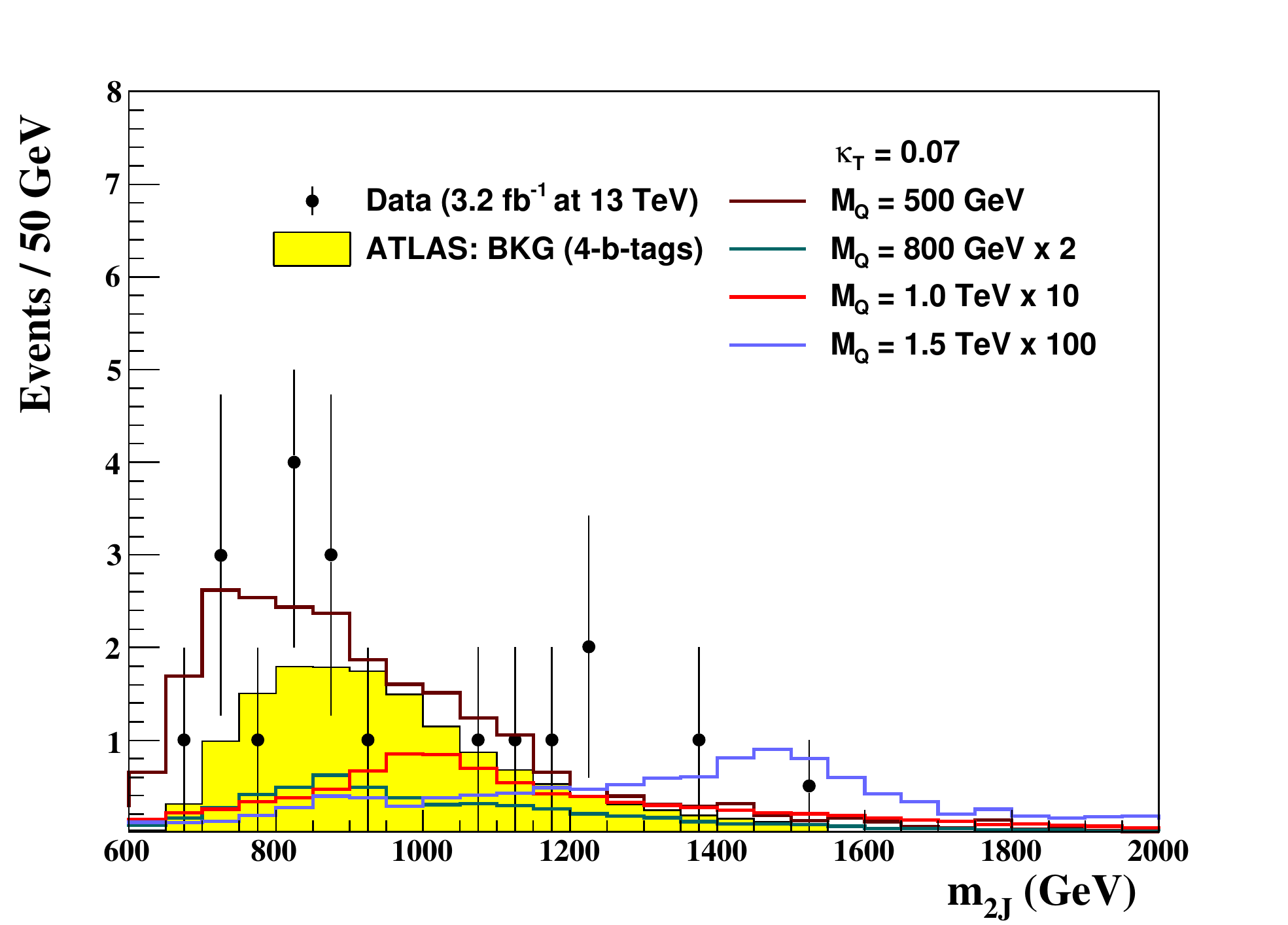}
  \includegraphics[width=0.48\textwidth]{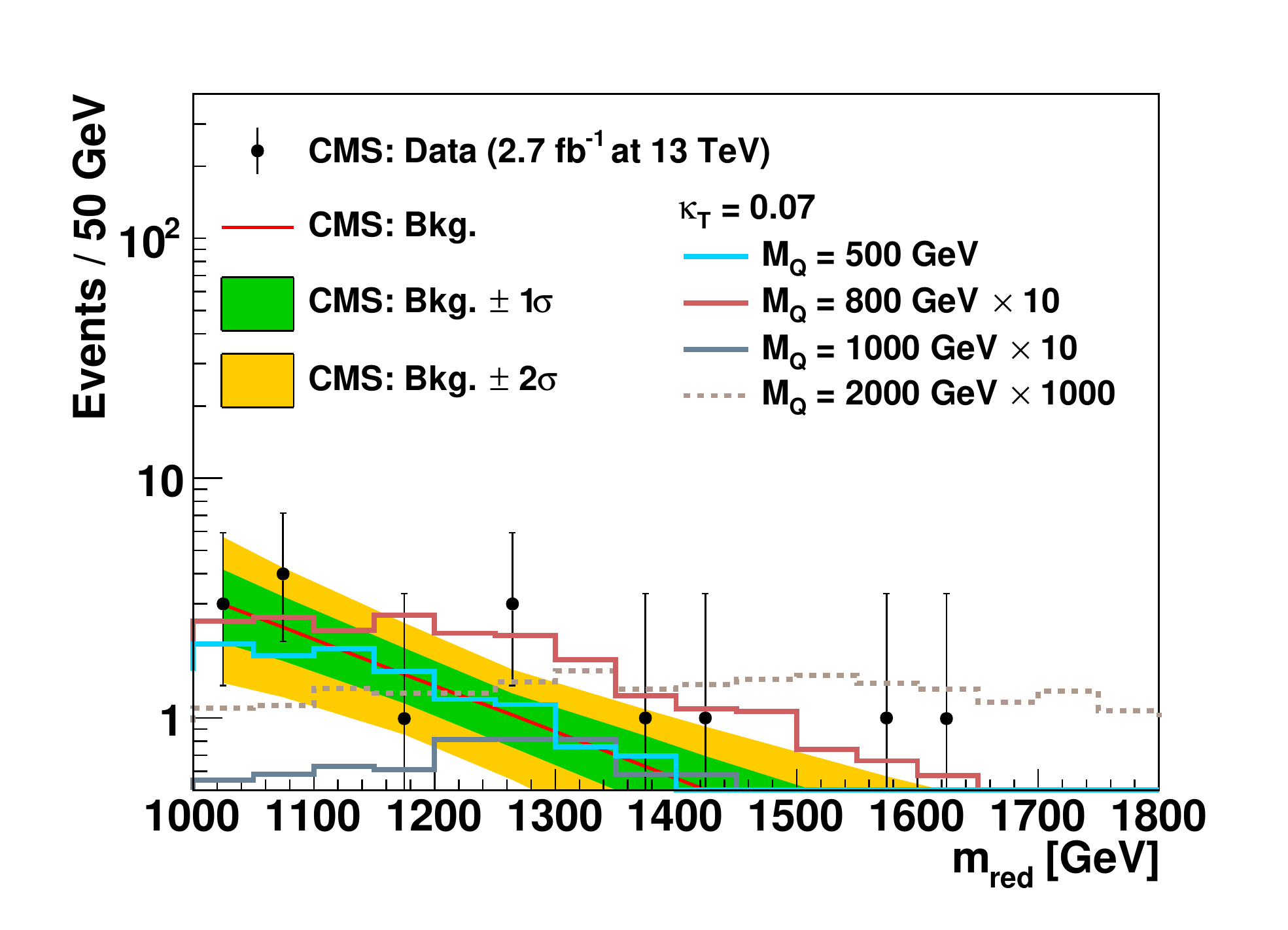}
  \caption{\footnotesize Distribution in the invariant mass of the system made
  of the two reconstructed boosted Higgs bosons after applying the boosted
  ATLAS-like analysis selection (left), and in the reduced mass resulting from
  the boosted CMS-like analysis (right). We select varied benchmark setups and
  overlay our predictions with the
  ATLAS~\cite{ Aaboud:2016xco} and CMS~\cite{CMS:2016pwo} data and backgrounds,
  respectively.}
 \label{fig:ATLAS_CMS_boostedRecast}
\end{figure}

The four analyses that we have reimplemented have not been designed to
target VLQ-induced di-Higgs production in the first place, as reflected in the
comparison
of the differential distributions in the analysis key observables with data
illustrated in Figure~\ref{fig:ATLAS_CMS_resolvedRecast} in the resolved case.
In this figure, we
overlay our predictions for the spectrum in the reconstructed di-Higgs
invariant-mass $m_{4j}$ (built from the four leading jets) with the
results obtained by the ATLAS (left panel) and CMS (right panel) collaborations
in their resolved di-Higgs analysis of 3.2~fb$^{-1}$ and 2.3~fb$^{-1}$ of
$13$\,TeV LHC data, respectively~\cite{Aaboud:2016xco,CMS:2016tlj}.
In Figure~\ref{fig:ATLAS_CMS_boostedRecast}, we focus on the boosted analysis
case and we respectively show the distribution in the
invariant mass of the system made of the two boosted Higgs bosons, $m_{2J}$, in
the case of the ATLAS-like boosted analysis (left panel) and the reduced mass
$m_{\rm red}$ obtained in the context of the boosted CMS-like analysis (right
panel) as defined in Eq.~\eqref{eq:mred}. In both cases, the theoretical
predictions and the data, respectively extracted from the ATLAS~\cite{%
ATLAS:2016ixk} and CMS~\cite{CMS:2016pwo} publications, are superimposed.

Although it may be challenging for a resolved analysis to be sensitive to
VLQ-induced di-Higgs production for large VLQ masses, boosted analyses in
principle offer extra
handles for extending the sensitivity up to the TeV scale. The official numbers
of data events and the predicted numbers of signal events for $\kappa_{\rm T} =
0.07$ are given in Table~\ref{tab:recastYields} for various VLQ masses.
Comparing the magnitude of the yields, it turns out that ATLAS and CMS are
already sensitive to a large fraction of the model parameter space with present
data. This prevents us from requiring a specific design of a VLQ-dedicated
di-Higgs search.

\begin{table}
\begin{center}
\setlength\tabcolsep{6pt}
\renewcommand{\arraystretch}{1.2}
\begin{tabular}{l||c|c||c|c|c|c}
  \multirow{2}{*}{Analysis} & \multirow{2}{*}{Data} &\multirow{2}{*}{SM} &
  \multicolumn{4}{c}{Signals (for given VLQ masses)} \\
  \cline{4-7} & & & 0.5~TeV & 0.8~TeV & 1~TeV & 2~TeV \\
  \hline
  \hline
  ATLAS-resolved (3.2~fb$^{-1}$) & 44  & $47.6\pm3.8$ & 47.0 & 3.34 & 0.78 & -     \\
  ATLAS-boosted  (3.2~fb$^{-1}$) & 20  & $14.6\pm2.4$ & 23.5 & 2.82 & 0.933 & 0.024 \\
  \hline
  CMS-resolved   (2.3~fb$^{-1}$) & 797 & n.a.         & 120  & 7.27 & 1.68 & -     \\ 
  CMS-boosted    (2.7~fb$^{-1}$) & 15  & n.a.         & 17.6 & 2.99 & 1.10 & 0.04  \\  
\end{tabular}
  \caption{Number of data and predicted signal and background events for four
  VLQ masses of
  $M_Q =$\, 500, 800, 1000, and 2000\,GeV, as obtained in the four reinterpreted
  LHC analyses and for $\kappa_{\rm T} = 0.07$.}\label{tab:recastYields}
\end{center}
\end{table}

\section{Characterising the signal\label{sec:signalcat}}

\begin{figure}
\centering
\includegraphics[width=0.32\textwidth]{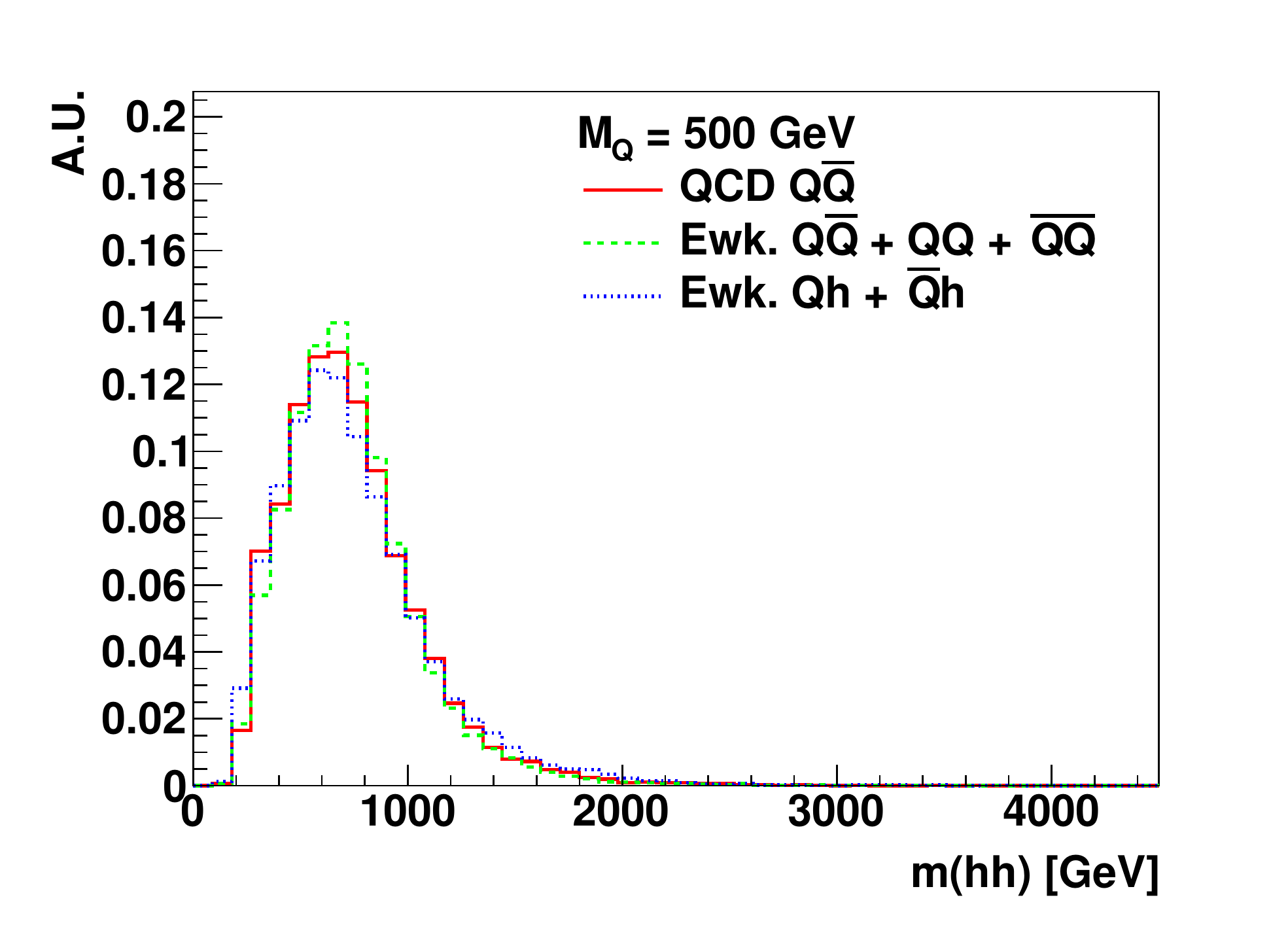}
\includegraphics[width=0.32\textwidth]{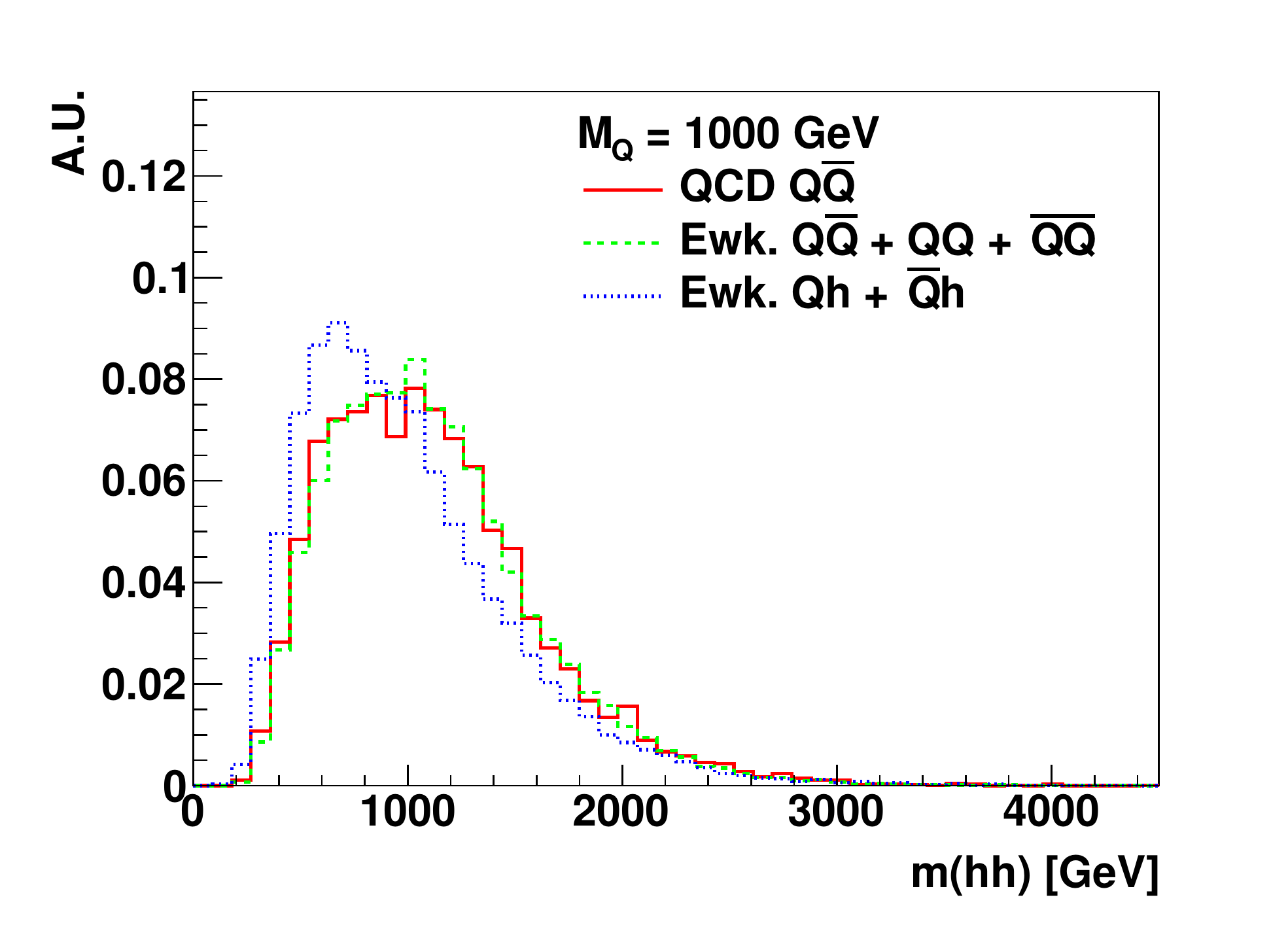}
\includegraphics[width=0.32\textwidth]{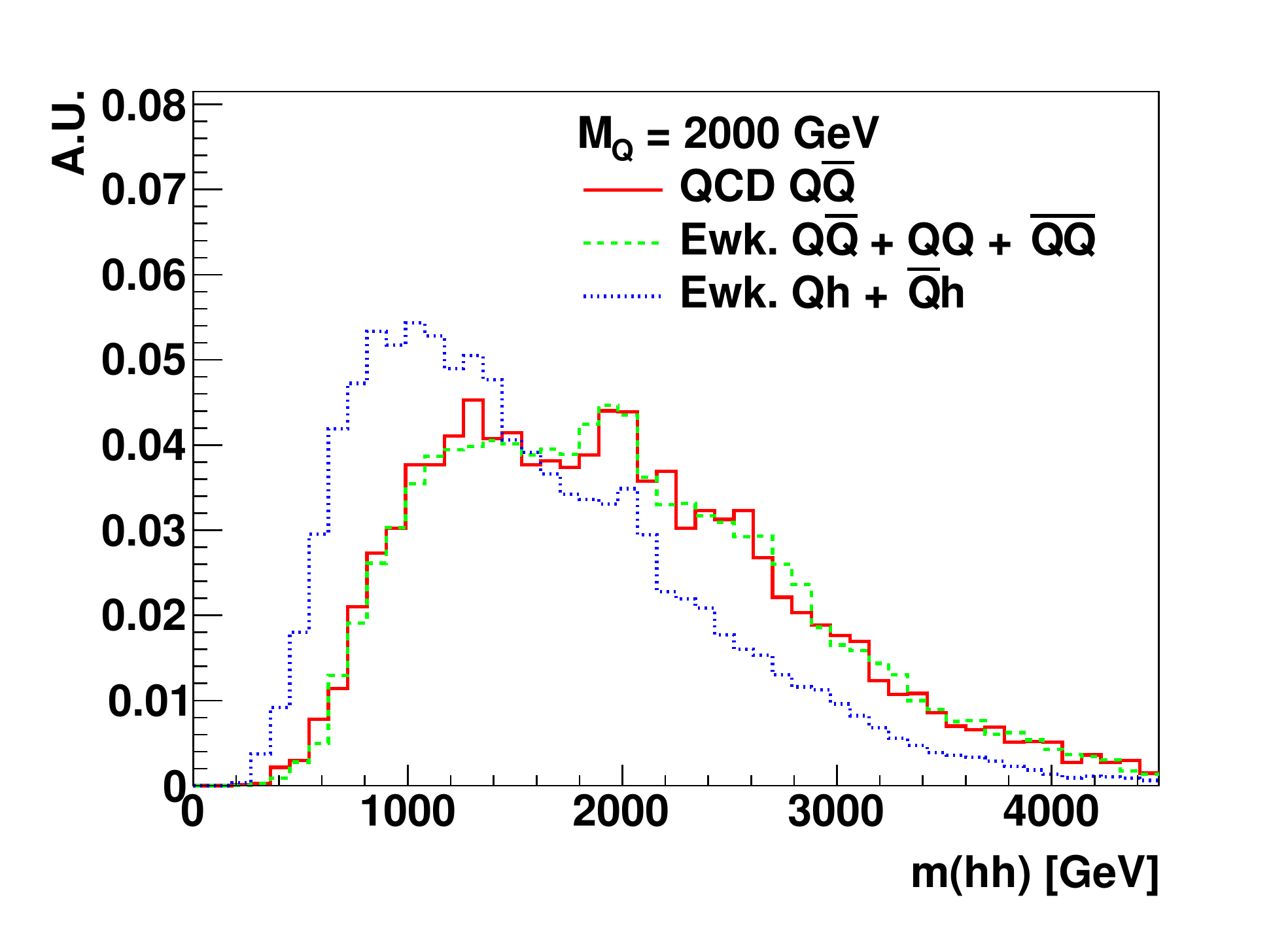}
  \caption{Normalised distribution, at the NLO accuracy in QCD, in the di-Higgs
  invariant mass in the context of VLQ-induced di-Higgs production. We
  distinguish QCD-induced (red) and EW-induced (green) VLQ-pair production
  (followed by to $Q\to hq$ decays) and the associated production of a
  single VLQ together with a Higgs boson (blue). We set the VLQ mass to 500~GeV
  (left panel), 1000~GeV (middle panel) and 2000~GeV (right panel).
  \label{fig:sigChar_MHH}}\vspace*{4mm}
\includegraphics[width=0.32\textwidth]{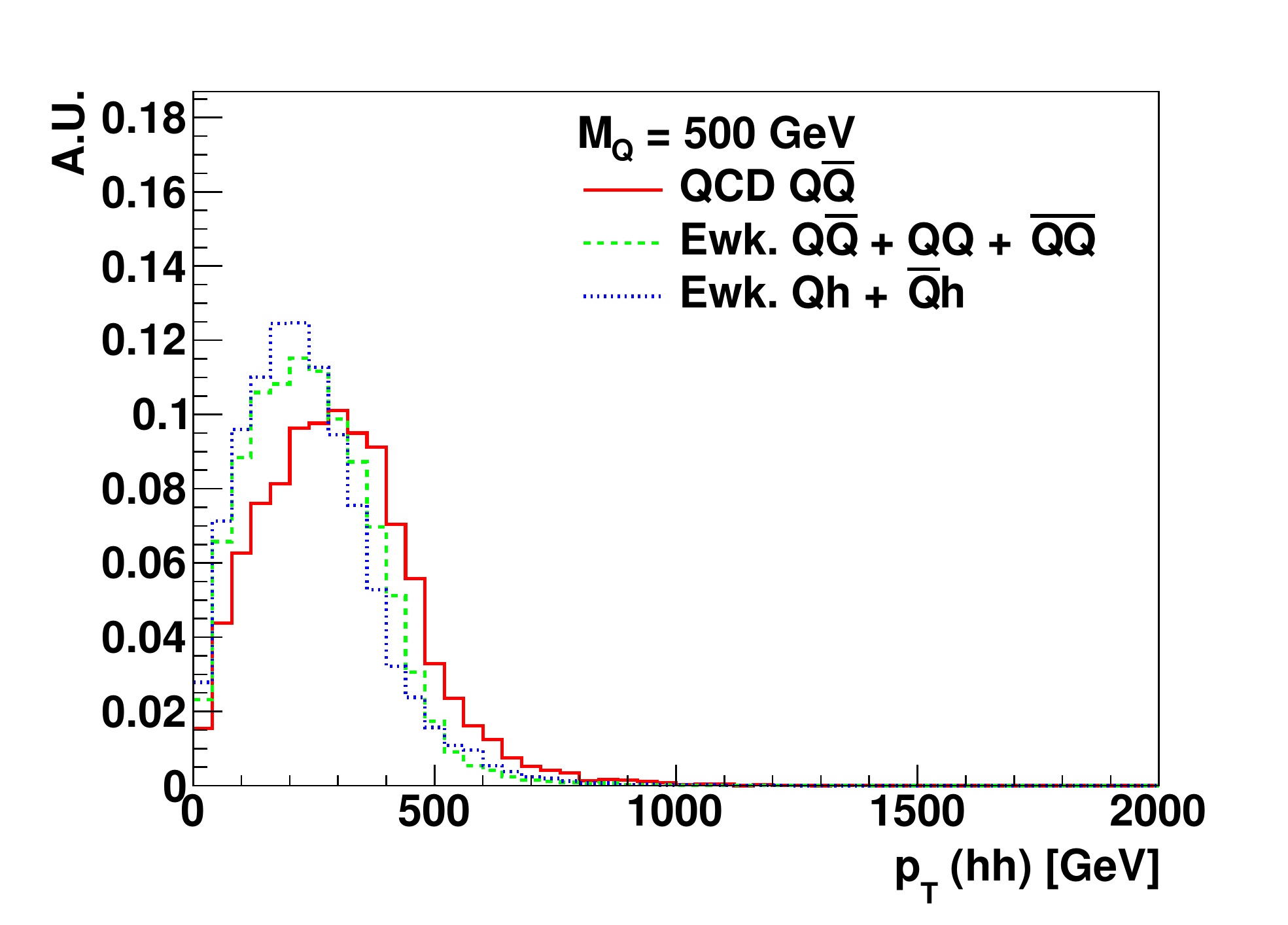}
\includegraphics[width=0.32\textwidth]{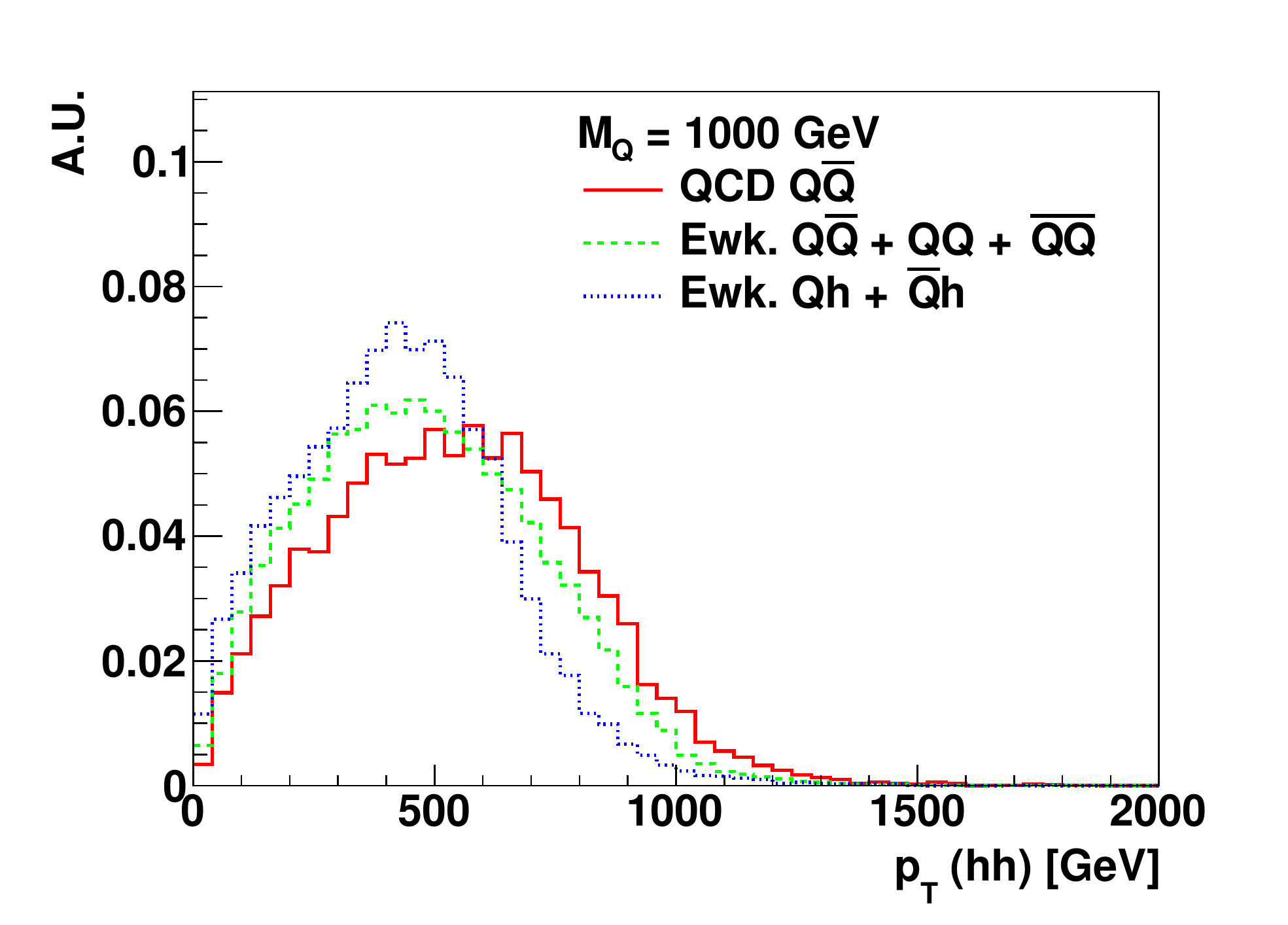}
\includegraphics[width=0.32\textwidth]{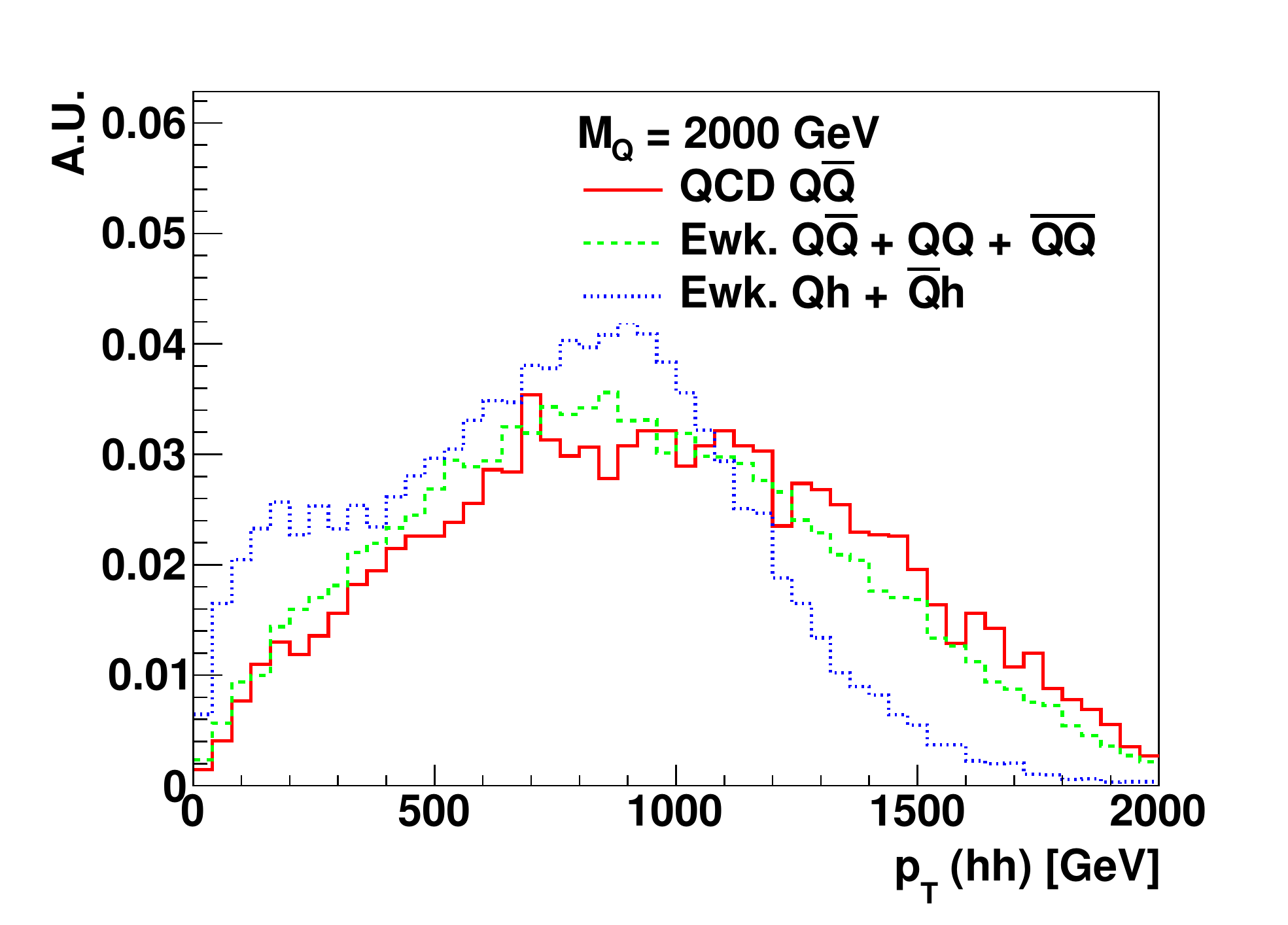}
  \caption{Same as Figure~\ref{fig:sigChar_MHH} but for the transverse momentum  of the
 di-Higgs system. \label{fig:sigChar_PTHH}}\vspace*{4mm}
\includegraphics[width=0.32\textwidth]{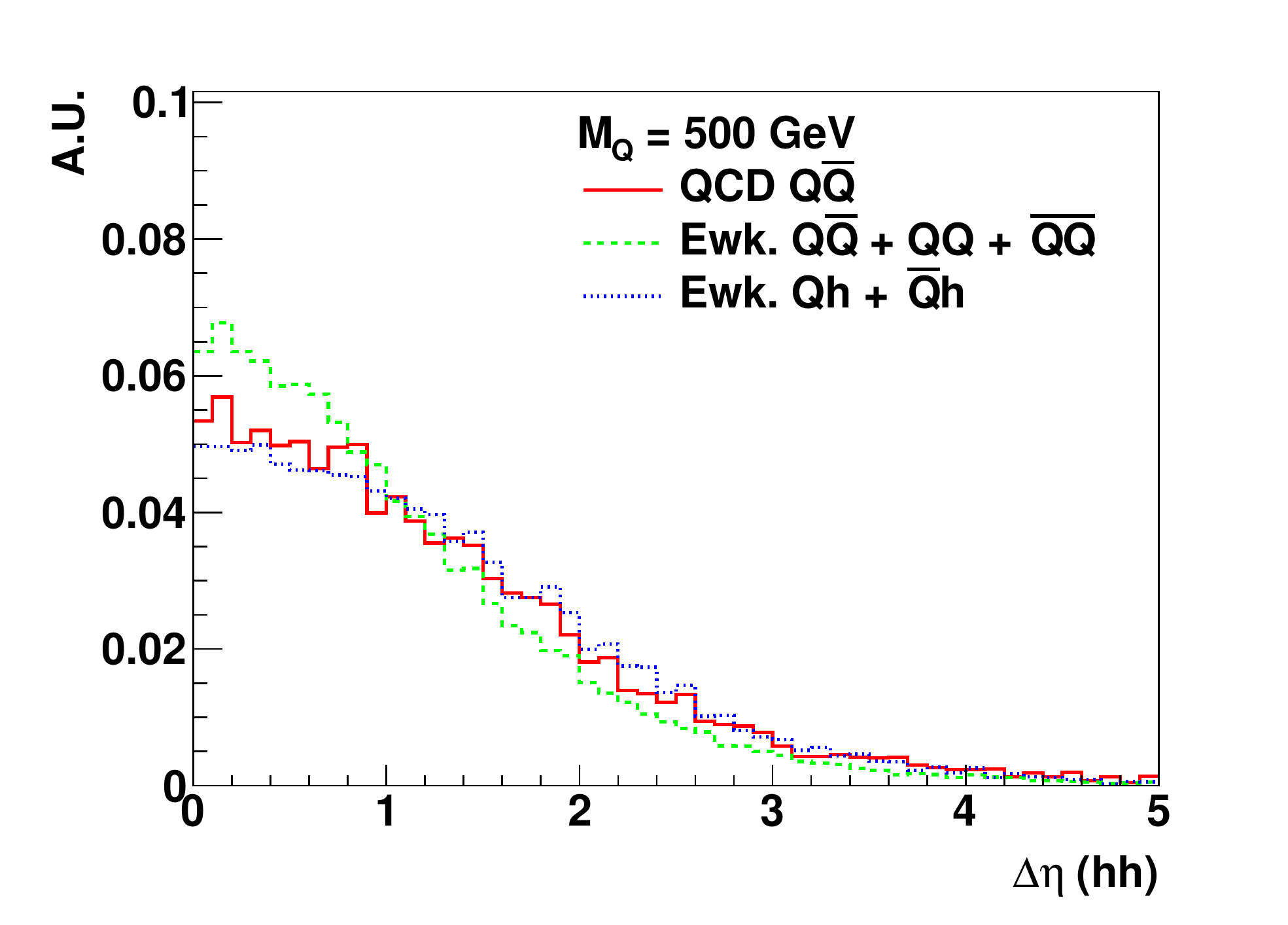}
\includegraphics[width=0.32\textwidth]{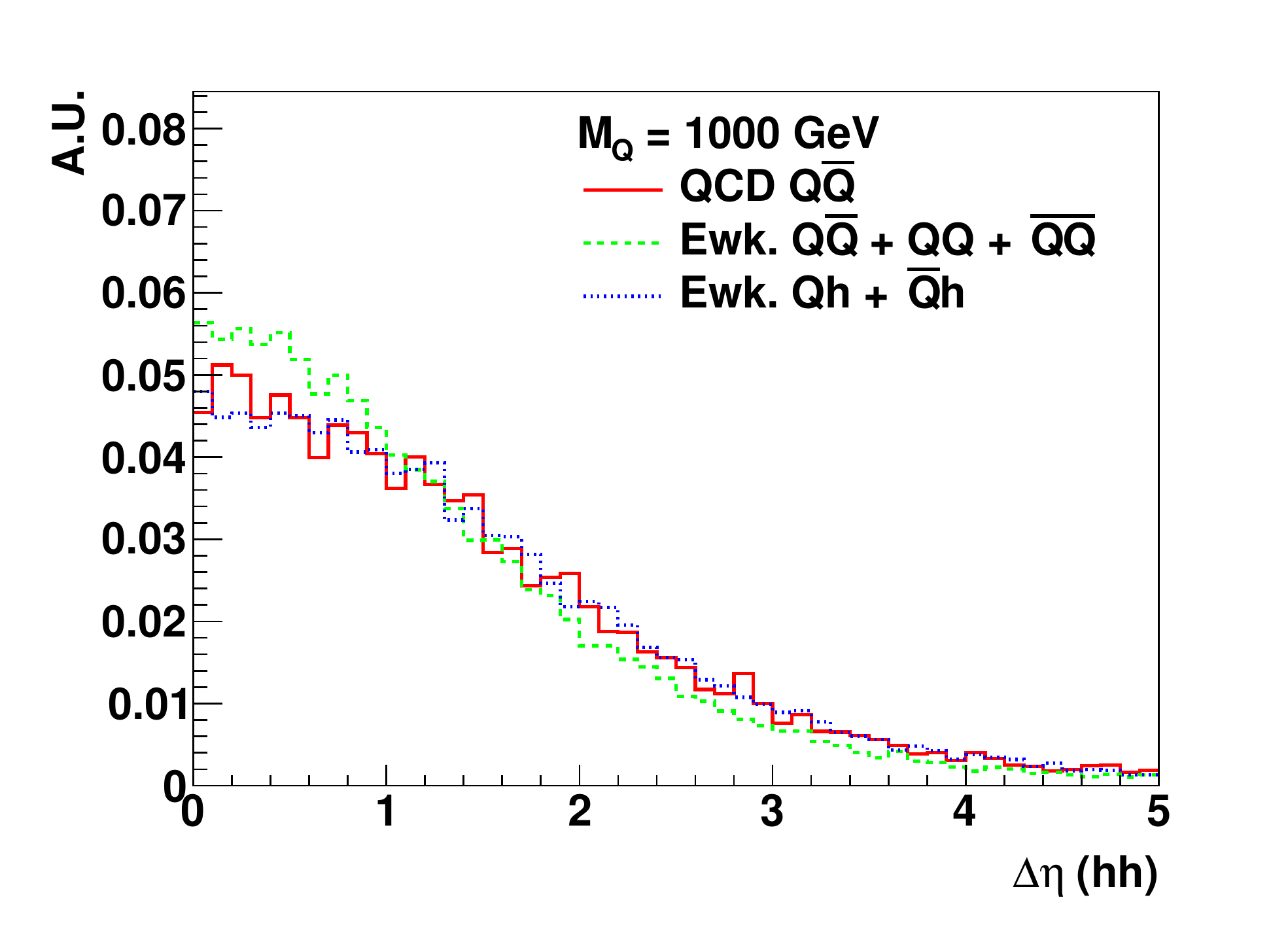}
\includegraphics[width=0.32\textwidth]{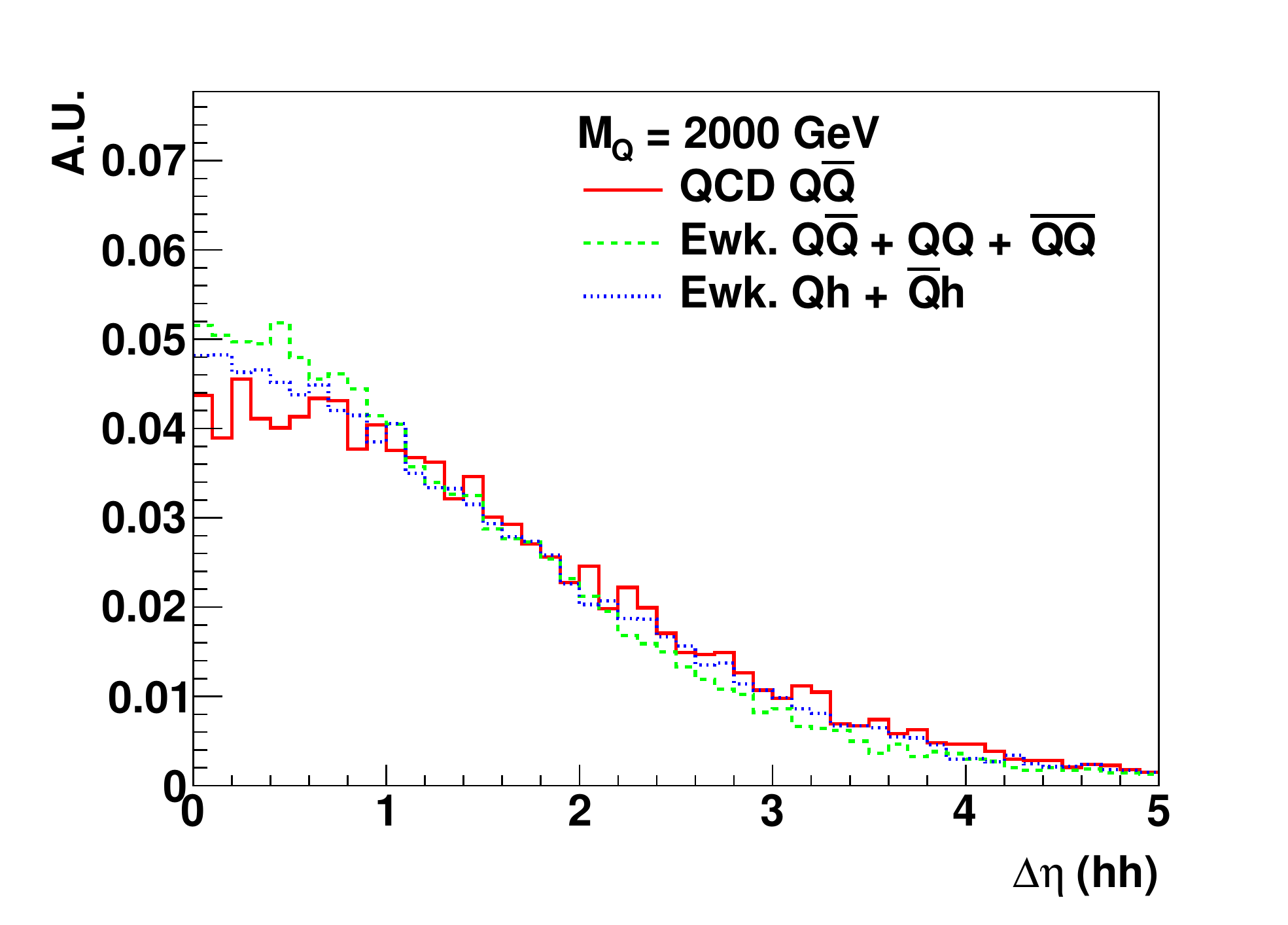}
\caption{Same as Figure~\ref{fig:sigChar_MHH} but for the separation in
  pseudorapidity of the di-Higgs system.
\label{fig:sigChar_DEtaHH}}
\end{figure}

In this section, we study properties of the VLQ-induced di-Higgs signal that
could be used to characterise it.
Whereas the QCD production of a VLQ-pair (and thus
of a di-Higgs system) is independent of $\kappa_Q$, the corresponding EW channel
rate scales like $\kappa_Q^4$, whilst the rate for the associated production of
a VLQ with a Higgs boson scales like $\kappa_Q^2$.

The knowledge of the resolved
or boosted regime is important, and we additionally consider a semi-boosted
context where only one of the two Higgs bosons is boosted. This new analysis
strategy is expected to be important in asymmetric cases where the transverse
momenta of the two Higgs bosons are largely different, as it could happen from
non-resonant Higgs-boson pair production. We therefore define, in this section,
three categories that we denote by {\it boosted}, {\it semi-boosted} and {\it
resolved}.

The fully boosted regime includes events where at least two Higgs fat jets are
found after following the boosted CMS-like analysis of
Section~\ref{ss:BoostedAna} with
the exception of the $\Delta\eta$ selection of Eq.~\eqref{eq:CMSDeta} that we
omit. We tag as semi-boosted events in which only one Higgs fat jet is found,
but that contains two additional $b$-jets that are consistent with a \Hbb decay
as defined in the resolved CMS-like analysis of Section~\ref{ss:ResolvedAna}. We
further impose the two reconstructed Higgs bosons to satisfy
Eq.~\eqref{eq:chi2CMS}.
Finally, the fully resolved category admits events with four well-identified
$b$-jets as in the resolved CMS-like selection of Section~\ref{ss:ResolvedAna}.

\begin{figure}
\centering
\includegraphics[width=0.32\textwidth]{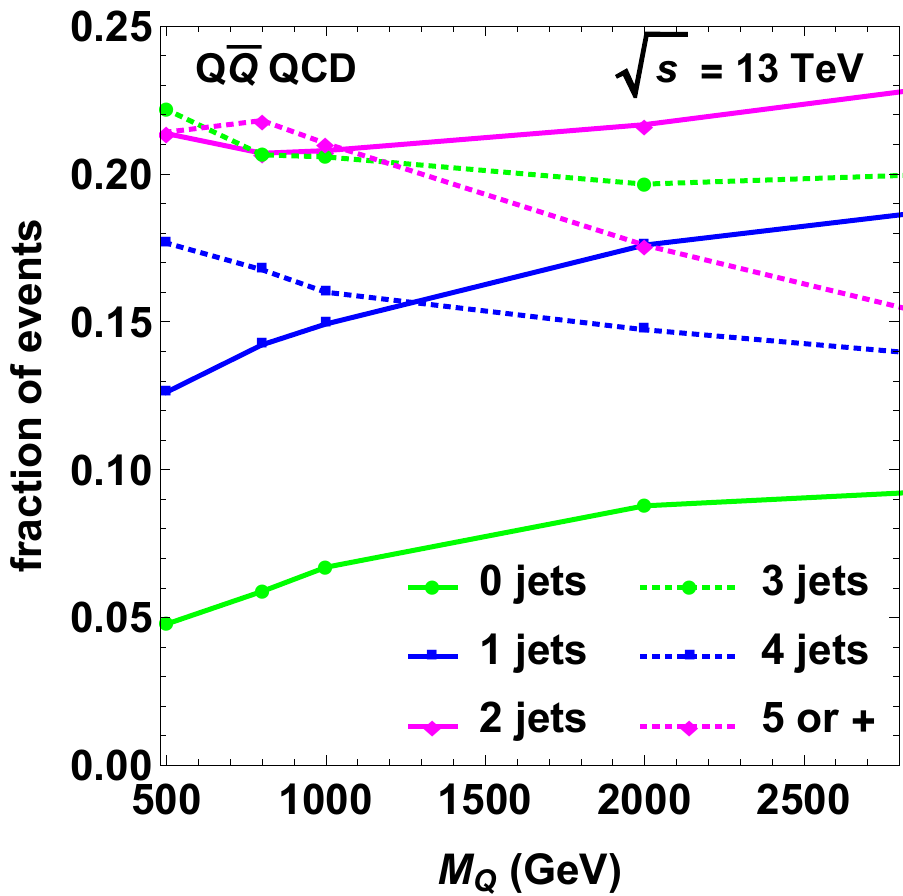}
\includegraphics[width=0.32\textwidth]{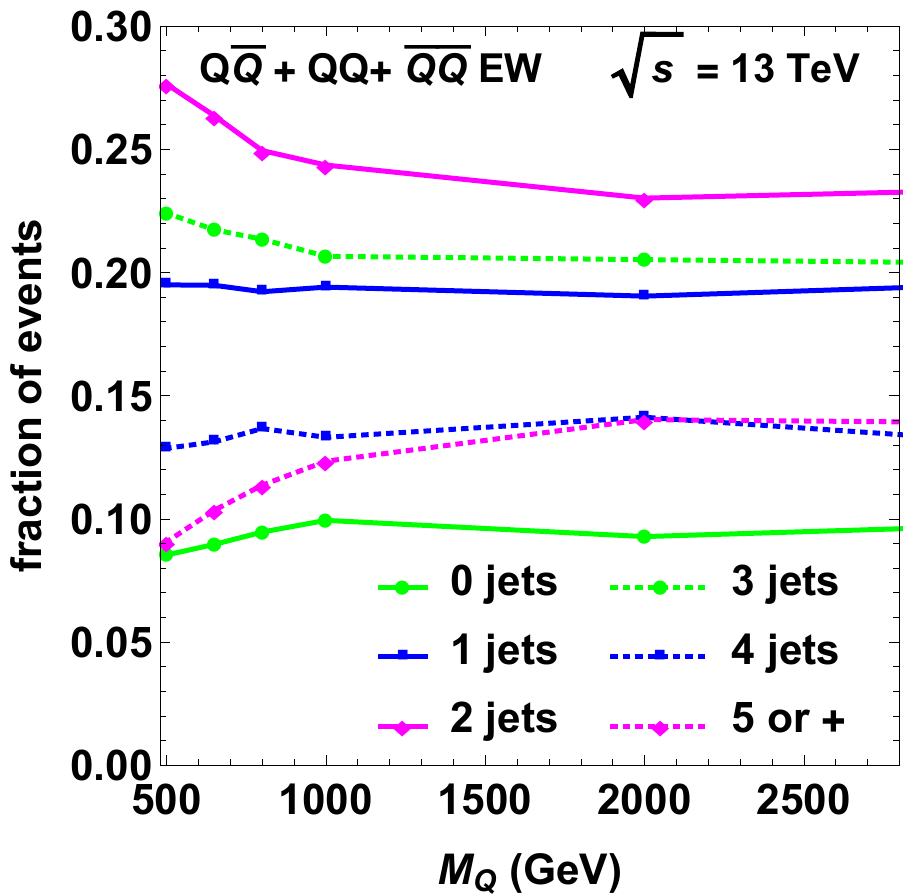}
\includegraphics[width=0.32\textwidth]{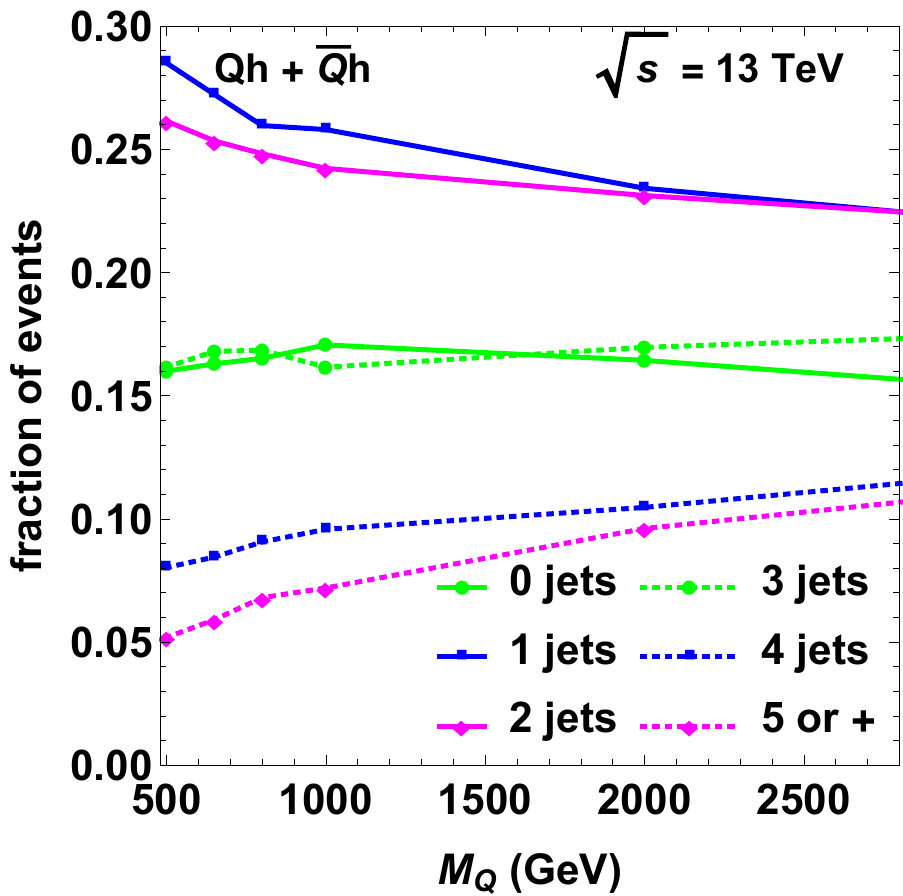}
\caption{Fraction of selected signal events featuring a given number of AK4 jets
  that are not originating from a Higgs-boson decay. We separately consider
  QCD-induced (left panel) and EW-induced (middle panel) VLQ pair production, as
  well as associated Higgs and VLQ production (right panel).}
\label{fig:cat_njets_proc_mass}
\end{figure}

We analyse the properties of the two reconstructed Higgs bosons for VLQ masses
of 500~GeV (left panel of the following figures), 1000~GeV (central panel of the
following figures) and 2000~GeV (right panel of the following figures). The
results are summed over all the three above-mentioned categories. In
Figure~\ref{fig:sigChar_MHH}, we present the
distribution in the invariant mass of the reconstructed di-Higgs system. The
spectra are indistinguishable for VLQ of mass equal to 500~GeV, as in this case
the two Higgs bosons are mostly produced at rest. When one increases $M_Q$, the
Higgs bosons start to become more boosted, which results in differences in the
spectrum that depend on the production mode, the peak appearing at 2~TeV being
related to a configuration where the two Higgs bosons are back-to-back.

More pronounced differences between the three different production channels can
be observed by studying other kinematic variables like the transverse momentum
of the di-Higgs system, while other observables exhibit very similar spectra. In
all cases, it will not be easy to use the information to disambiguate the
production modes. This is illustrated by Figure~\ref{fig:sigChar_PTHH} and
Figure~\ref{fig:sigChar_DEtaHH} where we respectively show the distribution in
the transverse momentum of the di-Higgs system and the separation in
pseudorapidity between the two Higgs bosons.
The $p_T$ of the Higgs boson pair has a harder shape for the
QCD pair production channel, while the difference in pseudorapidity between the
two Higgs bosons features a softer shape for EW VLQ pair production.
The jet properties are investigated in
Figure~\ref{fig:cat_njets_proc_mass} where we investigate the AK4 jet
multiplicity this time
for the signal selected events. We present the fraction of selected events
featuring $n$ jets with $n$ ranging from 0 to at least 5 jets as a function of
the VLQ mass. Selected signal events arising from VLQ (QCD and EW) pair
production feature most of the time between one and three extra jets, whereas
di-Higgs production via a $Qh$ pair in general leads to one or two extra jets
only. The difference is however rather mild, so that jet vetoes would be poor
handles to separate the different components of the signal.

\begin{figure}
\centering
\includegraphics[width=0.32\textwidth]{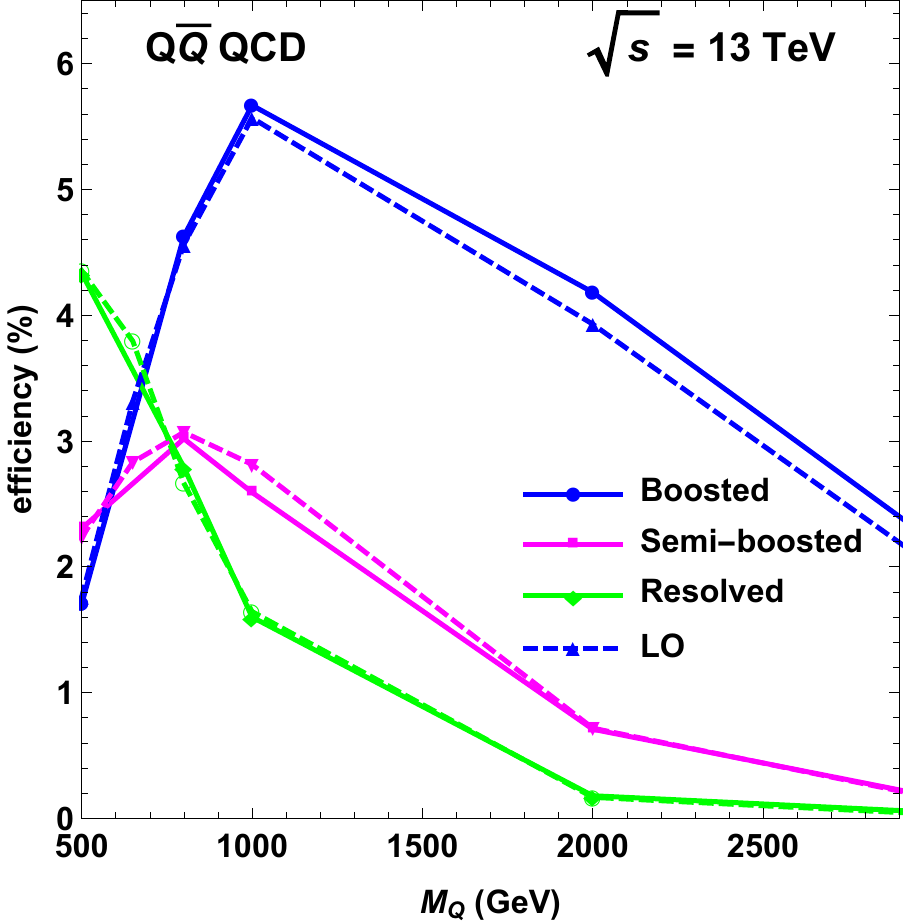}
\includegraphics[width=0.32\textwidth]{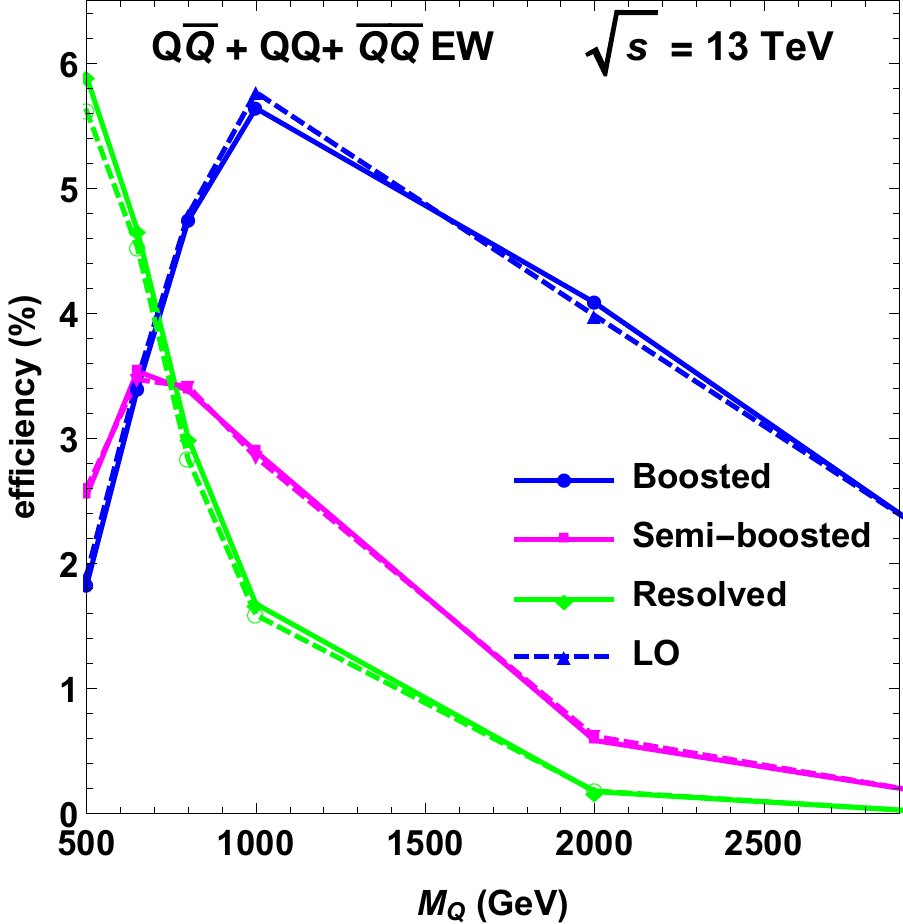}
\includegraphics[width=0.32\textwidth]{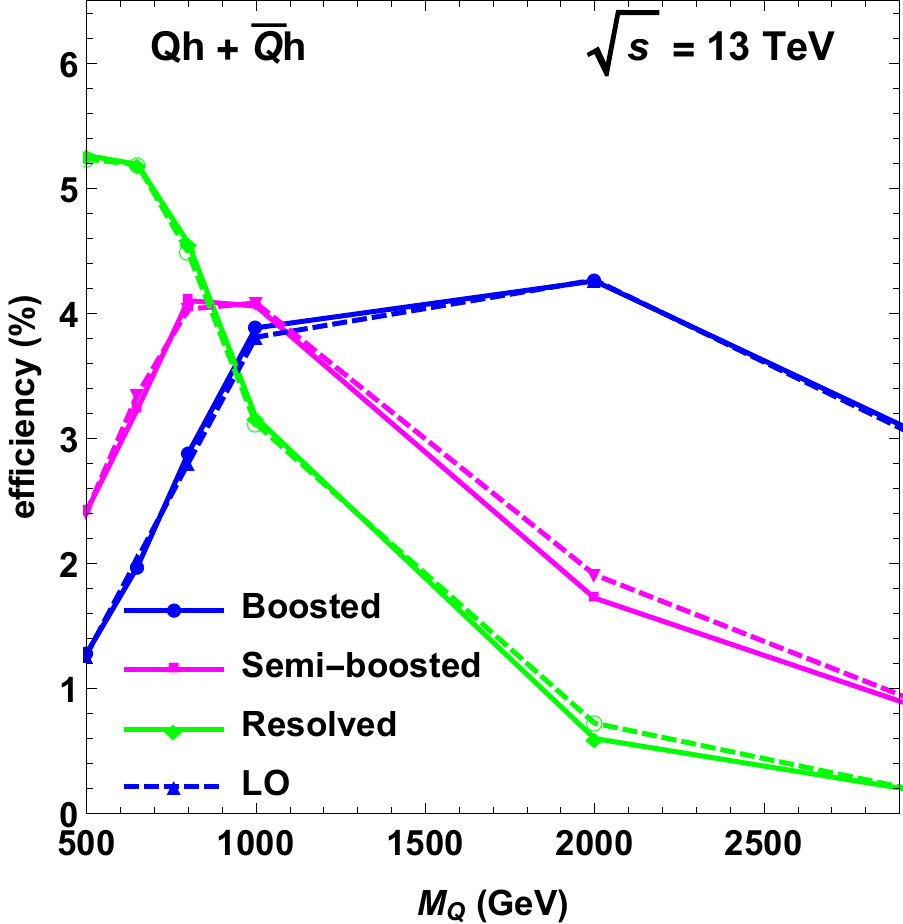}
\caption{\footnotesize{
 Signal selection efficiencies as a function of the VLQ mass for the boosted
 (blue), semi-boosted (magenta) and resolved (green)
 analysis categories and separated in terms of the different components of the
 signal, namely QCD (left panel) and EW (middle panel) VLQ-induced production
 and $Qh$ (right panel) production. The solid lines refer to NLO simulations
 while the dashed lines refer to a LO simulation.
\label{fig:cat_mjj_proc_mass}}}
\end{figure}

Finally, we present in Figure~\ref{fig:cat_mjj_proc_mass} the signal
efficiencies as a function of the VLQ mass for the different analysis categories
and the different components of the signal, and assess the impact of using
NLO-accurate simulations (solid lines) instead LO-accurate simulations (dashed
lines). While NLO effects are in general mild at the level of the efficiency for
the EW processes (so that only the total rate turns to be modified), they are
noticeable for the QCD production channel and can reach about 10\%.

The results
indicate that designing
a semi-boosted category is useful as many events feature a single boosted Higgs
boson and not a pair of them, in particular for VLQ masses of about 1~TeV. This
effect is as expected more pronounced for $Qh$ production where the two Higgs
bosons are produced in an asymmetric fashion, one of them being directly
produced and the
second of them being originating from a VLQ decay. On different lines, the
efficiency of the resolved analysis for what concerns EW VLQ-pair-induced
di-Higgs production is very large in the low mass region, so that the inclusion
of this channel can be really helpful to further constrain the low-mass region
of the parameter space when the $\kappa_Q$ coupling strength is not too small.

Our phenomenological analysis has moreover shown that the three different
production channels have different kinematic properties so that further
selection cuts could be implemented in the aim of distinguishing them once a
signal is observed at the LHC. While no distribution seems sufficient by
itself, the usage of a combination of all the available pieces of information
simultaneously, via, {\it e.g.}, a multivariate analysis technique, could allow
to achieve a sufficiently large discriminating power. Other channels and
variables could nonetheless be useful for characterising any potential excess
and disentangling the theoretical setups that could accommodate the
excess~\cite{Backovic:2014ega}.

\section{Conclusions} 
\label{sec:con}
Many extensions of the Standard Model feature one or more than one vector-like
quark multiplets. The presence of several extra multiplets of quarks is
motivated both by their impact on the predictions for various well-measured
observables, which allows one to ease the corresponding experimental
constraints, and by theoretical considerations. For instance, the combination of
extra quark multiplets could tame down new physics effects on the
couplings of the $Z$-boson to quarks, or could originate from the symmetry
structure of a model where the custodial symmetry is enforced. In such
scenarios, the model may accommodate vector-like quarks in the 
low-energy part of the particle spectrum that dominantly couple
to the Higgs boson and the SM quarks.

We have provided in the present study working examples of such a situation and
developed a model-independent implementation in terms of an effective Lagrangian
with parameters related to the physical observables, and we have connected this
Lagrangian to two realistic ultraviolet-complete setups. We have further
investigated the potential impact of di-Higgs probes for discovering or
constraining the model, and we have detailed the variety of channels giving rise
to a di-Higgs signature that could be mediated by (single and pair produced)
vector-like quarks. Our predictions include next-to-leading order corrections in
QCD and show their relevance not only in terms of cross-section values, but also
for the increased precision with respect to the uncertainties.

We have furthermore considered the decay of the two Higgs bosons into four $b$
quarks, and performed simulations corresponding in a close way to the existing
Run II analyses performed by ATLAS and CMS both in the so-called resolved and
boosted regimes. We have looked in details to the resulting signal efficiencies
and to kinematical observables and their usage as handles on new physics. The
current experimental analyses are mostly based on the study of resonant
channels, which
are not optimal in the case of a di-Higgs signal stemming from the interactions
of vector-like quarks with the Higgs boson. Nonetheless, the Higgs bosons are
typically sufficiently boosted to yield acceptable efficiencies for VLQ masses
larger than 500 GeV.

Our study also allows to go to a further level of characterisation of a
potential signal, using specific kinematical observables which can discriminate
the VLQ-induced di-Higgs production mechanisms. The conclusion of such a study
is that there is an opportunity of improvement with respect to the present
bounds (or discovery) in the forthcoming LHC analyses, under the condition that
a study tailored on the VLQ-induced di-Higgs channels is performed.

\section*{Acknowledgements}

We thank the organisers of the 2015 ‘{\it Les Houches – Physics at TeV
colliders}’
workshop where this work was initiated. GC, HC and AD acknowledge partial
support from the Labex-LIO (Lyon Institute of Origins) under grant
ANR-10-LABX-66 and FRAMA (FR3127, F\'ed\'eration de Recherche `Andr\'e Marie
Amp\`ere'). The work of HSS and BF is supported in part by French state
funds managed by the Agence Nationale de la Recherche (ANR), in the
context of the LABEX ILP (ANR-11-IDEX-0004-02, ANR-10-LABX-63), and the one of
TF by the Basic Science Research Program through the National Research
Foundation of Korea (NRF) funded by the ministry of Education, Science and
Technology (No.~2013R1A1A1062597) and by IBS under the project code
IBS-R018-D1. AC has been supported by the MIURFIRB RBFR12H1MW grant.
GC, HC, AD, BF and TF also acknowledge support from the France
Korea Particle Physics LIA (FKPPL) and from the CNRS `Partenariat Hubert-Curien'
(PHC) STAR project number 34299VE. 
%DM is supported by the National Science Foundation under award PHY-1306953. 
DM is supported by the National Science Foundation under award PHY-1607262. 

\bibliography{VLQ}

\providecommand{\href}[2]{#2}\begingroup\raggedright\begin{thebibliography}{10}

\bibitem{Antoniadis:1990ew}
I.~Antoniadis, \emph{{A Possible new dimension at a few TeV}},
  \href{http://dx.doi.org/10.1016/0370-2693(90)90617-F}{\emph{Phys. Lett. B}
  {\bf 246} (1990) 377--384}.

\bibitem{Kaplan:1991dc}
D.~B. Kaplan, \emph{{Flavor at SSC energies: A New mechanism for dynamically
  generated fermion masses}},
  \href{http://dx.doi.org/10.1016/S0550-3213(05)80021-5}{\emph{Nucl. Phys. B}
  {\bf 365} (1991) 259--278}.

\bibitem{ArkaniHamed:2002qx}
N.~Arkani-Hamed, A.~G. Cohen, E.~Katz, A.~E. Nelson, T.~Gregoire and J.~G.
  Wacker, \emph{{The Minimal moose for a little Higgs}},
  \href{http://dx.doi.org/10.1088/1126-6708/2002/08/021}{\emph{JHEP} {\bf 08}
  (2002) 021}, [\href{http://arxiv.org/abs/hep-ph/0206020}{{\tt
  hep-ph/0206020}}].

\bibitem{Contino:2006qr}
R.~Contino, L.~Da~Rold and A.~Pomarol, \emph{{Light custodians in natural
  composite Higgs models}},
  \href{http://dx.doi.org/10.1103/PhysRevD.75.055014}{\emph{Phys. Rev. D} {\bf
  75} (2007) 055014}, [\href{http://arxiv.org/abs/hep-ph/0612048}{{\tt
  hep-ph/0612048}}].

\bibitem{Matsedonskyi:2012ym}
O.~Matsedonskyi, G.~Panico and A.~Wulzer, \emph{{Light Top Partners for a Light
  Composite Higgs}},
  \href{http://dx.doi.org/10.1007/JHEP01(2013)164}{\emph{JHEP} {\bf 01} (2013)
  164}, [\href{http://arxiv.org/abs/1204.6333}{{\tt 1204.6333}}].

\bibitem{Atre:2013ap}
A.~Atre, M.~Chala and J.~Santiago, \emph{{Searches for New Vector Like Quarks:
  Higgs Channels}},
  \href{http://dx.doi.org/10.1007/JHEP05(2013)099}{\emph{JHEP} {\bf 05} (2013)
  099}, [\href{http://arxiv.org/abs/1302.0270}{{\tt 1302.0270}}].

\bibitem{Azatov:2012rj}
A.~Azatov, O.~Bondu, A.~Falkowski, M.~Felcini, S.~Gascon-Shotkin, D.~K. Ghosh
  et~al., \emph{{Higgs boson production via vector-like top-partner decays:
  Diphoton or multilepton plus multijets channels at the LHC}},
  \href{http://dx.doi.org/10.1103/PhysRevD.85.115022}{\emph{Phys. Rev. D} {\bf
  85} (2012) 115022}, [\href{http://arxiv.org/abs/1204.0455}{{\tt 1204.0455}}].

\bibitem{ATLAS:2016ixk}
{\scshape ATLAS} collaboration, \emph{{Search for pair production of Higgs
  bosons in the $b\bar{b}b\bar{b}$ final state using proton$-$proton collisions
  at $\sqrt{s} = 13$ TeV with the ATLAS detector}},  ATLAS-CONF-2016-049.

\bibitem{Aaboud:2016xco}
{\scshape ATLAS} collaboration, M.~Aaboud et~al., \emph{{Search for pair
  production of Higgs bosons in the $b\bar{b}b\bar{b}$ final state using
  proton--proton collisions at $\sqrt{s} = 13$ TeV with the ATLAS detector}},
  \href{http://dx.doi.org/10.1103/PhysRevD.94.052002}{\emph{Phys. Rev.} {\bf
  D94} (2016) 052002}, [\href{http://arxiv.org/abs/1606.04782}{{\tt
  1606.04782}}].

\bibitem{CMS:2016foy}
{\scshape CMS} collaboration, \emph{{Search for non-resonant pair production of
  Higgs bosons in the $\rm{b} \bar{\rm{b}} \rm{b} \bar{\rm{b}}$ final state
  with 13 TeV CMS data}},  CMS-PAS-HIG-16-026.

\bibitem{CMS:2016tlj}
{\scshape CMS} collaboration, \emph{{Search for resonant pair production of
  Higgs bosons decaying to two bottom quark-antiquark pairs in proton-proton
  collisions at 13 TeV}},  CMS-PAS-HIG-16-002.

\bibitem{CMS:2016pwo}
{\scshape CMS} collaboration, \emph{{Search for heavy resonances decaying to a
  pair of Higgs bosons in four b quark final state in proton-proton collisions
  at sqrt(s)=13 TeV}},  CMS-PAS-B2G-16-008.

\bibitem{CMS:2016knm}
{\scshape CMS} collaboration, \emph{{Search for resonant Higgs boson pair
  production in the $\mathrm{b\overline{b}}\tau^+\tau^-$ final state using 2016
  data}},  CMS-PAS-HIG-16-029.

\bibitem{CMS:2016ymn}
{\scshape CMS} collaboration, \emph{{Search for non-resonant Higgs boson pair
  production in the bbtautau final state using 2016 data}},
  CMS-PAS-HIG-16-028.

\bibitem{ATLAS:2016lala}
{\scshape ATLAS} collaboration, \emph{{Search for Higgs boson pair production
  in the $b\bar{b}\gamma\gamma$ final state using pp collision data at
  $\sqrt{s}=13$ TeV with the ATLAS detector}},  ATLAS-CONF-2016-004.

\bibitem{CMS:2016vpz}
{\scshape CMS} collaboration, \emph{{Search for $H(bb)H(\gamma\gamma)$ decays
  at 13~TeV}},  CMS-PAS-HIG-16-032.

\bibitem{CMS:2016cdj}
{\scshape CMS} collaboration, \emph{{Search for Higgs boson pair production in
  the $\mathrm{b}\overline{\mathrm{b}} \mathrm{l}\nu \mathrm{l}\nu$ final state
  at $\sqrt{s} = 13~\mathrm{TeV}$}},  CMS-PAS-HIG-16-024.

\bibitem{ATLAS:2016qmt}
{\scshape ATLAS} collaboration, \emph{{Search for Higgs boson pair production
  in the final state of $\gamma\gamma WW^*$($\rightarrow l\nu jj$) using 13.3
  fb$^{-1}$ of $pp$ collision data recorded at $\sqrt{s}= $ 13 TeV with the
  ATLAS detector}},  ATLAS-CONF-2016-071.

\bibitem{Brooijmans:2016vro}
G.~Brooijmans et~al., \emph{{Les Houches 2015: Physics at TeV colliders - new
  physics working group report}},  in \emph{{9th Les Houches Workshop on
  Physics at TeV Colliders (PhysTeV 2015) Les Houches, France, June 1-19,
  2015}}, 2016.
\newblock \href{http://arxiv.org/abs/1605.02684}{{\tt 1605.02684}}.

\bibitem{Fuks:2016ftf}
B.~Fuks and H.-S. Shao, \emph{{QCD next-to-leading-order predictions matched to
  parton showers for vector-like quark models}},
  \href{http://dx.doi.org/10.1140/epjc/s10052-017-4686-z}{\emph{Eur. Phys. J.}
  {\bf C77} (2017) 135}, [\href{http://arxiv.org/abs/1610.04622}{{\tt
  1610.04622}}].

\bibitem{Buchkremer:2013bha}
M.~Buchkremer, G.~Cacciapaglia, A.~Deandrea and L.~Panizzi, \emph{{Model
  Independent Framework for Searches of Top Partners}},
  \href{http://dx.doi.org/10.1016/j.nuclphysb.2013.08.010}{\emph{Nucl. Phys. B}
  {\bf 876} (2013) 376--417}, [\href{http://arxiv.org/abs/1305.4172}{{\tt
  1305.4172}}].

\bibitem{Cacciapaglia:2011fx}
G.~Cacciapaglia, A.~Deandrea, L.~Panizzi, N.~Gaur, D.~Harada and Y.~Okada,
  \emph{{Heavy Vector-like Top Partners at the LHC and flavour constraints}},
  \href{http://dx.doi.org/10.1007/JHEP03(2012)070}{\emph{JHEP} {\bf 03} (2012)
  070}, [\href{http://arxiv.org/abs/1108.6329}{{\tt 1108.6329}}].

\bibitem{Ellis:2014dza}
S.~A.~R. Ellis, R.~M. Godbole, S.~Gopalakrishna and J.~D. Wells, \emph{{Survey
  of vector-like fermion extensions of the Standard Model and their
  phenomenological implications}},
  \href{http://dx.doi.org/10.1007/JHEP09(2014)130}{\emph{JHEP} {\bf 09} (2014)
  130}, [\href{http://arxiv.org/abs/1404.4398}{{\tt 1404.4398}}].

\bibitem{Cacciapaglia:2015ixa}
G.~Cacciapaglia, A.~Deandrea, N.~Gaur, D.~Harada, Y.~Okada and L.~Panizzi,
  \emph{{Interplay of vector-like top partner multiplets in a realistic mixing
  set-up}}, \href{http://dx.doi.org/10.1007/JHEP09(2015)012}{\emph{JHEP} {\bf
  09} (2015) 012}, [\href{http://arxiv.org/abs/1502.00370}{{\tt 1502.00370}}].

\bibitem{Ishiwata:2015cga}
K.~Ishiwata, Z.~Ligeti and M.~B. Wise, \emph{{New Vector-Like Fermions and
  Flavor Physics}},
  \href{http://dx.doi.org/10.1007/JHEP10(2015)027}{\emph{JHEP} {\bf 10} (2015)
  027}, [\href{http://arxiv.org/abs/1506.03484}{{\tt 1506.03484}}].

\bibitem{Dawson:2012di}
S.~Dawson and E.~Furlan, \emph{{A Higgs Conundrum with Vector Fermions}},
  \href{http://dx.doi.org/10.1103/PhysRevD.86.015021}{\emph{Phys. Rev.} {\bf
  D86} (2012) 015021}, [\href{http://arxiv.org/abs/1205.4733}{{\tt
  1205.4733}}].

\bibitem{Bizot:2015zaa}
N.~Bizot and M.~Frigerio, \emph{{Fermionic extensions of the Standard Model in
  light of the Higgs couplings}},
  \href{http://dx.doi.org/10.1007/JHEP01(2016)036}{\emph{JHEP} {\bf 01} (2016)
  036}, [\href{http://arxiv.org/abs/1508.01645}{{\tt 1508.01645}}].

\bibitem{Atre:2008iu}
A.~Atre, M.~Carena, T.~Han and J.~Santiago, \emph{{Heavy Quarks Above the Top
  at the Tevatron}},
  \href{http://dx.doi.org/10.1103/PhysRevD.79.054018}{\emph{Phys. Rev. D} {\bf
  79} (2009) 054018}, [\href{http://arxiv.org/abs/0806.3966}{{\tt 0806.3966}}].

\bibitem{Atre:2011ae}
A.~Atre, G.~Azuelos, M.~Carena, T.~Han, E.~Ozcan, J.~Santiago et~al.,
  \emph{{Model-Independent Searches for New Quarks at the LHC}},
  \href{http://dx.doi.org/10.1007/JHEP08(2011)080}{\emph{JHEP} {\bf 08} (2011)
  080}, [\href{http://arxiv.org/abs/1102.1987}{{\tt 1102.1987}}].

\bibitem{CMS:2016pul}
{\scshape CMS} collaboration, \emph{{Search for exotic light-flavor quark
  partners in pp collisions at $\sqrt{s} = 8~\mathrm{TeV}$}},
  CMS-PAS-B2G-12-016.

\bibitem{Agashe:2006at}
K.~Agashe, R.~Contino, L.~Da~Rold and A.~Pomarol, \emph{{A Custodial symmetry
  for $Zb \bar b$}},
  \href{http://dx.doi.org/10.1016/j.physletb.2006.08.005}{\emph{Phys. Lett. B}
  {\bf 641} (2006) 62--66}, [\href{http://arxiv.org/abs/hep-ph/0605341}{{\tt
  hep-ph/0605341}}].

\bibitem{Delaunay:2013pwa}
C.~Delaunay, T.~Flacke, J.~Gonzalez-Fraile, S.~J. Lee, G.~Panico and G.~Perez,
  \emph{{Light Non-degenerate Composite Partners at the LHC}},
  \href{http://dx.doi.org/10.1007/JHEP02(2014)055}{\emph{JHEP} {\bf 02} (2014)
  055}, [\href{http://arxiv.org/abs/1311.2072}{{\tt 1311.2072}}].

\bibitem{Dib:2005re}
C.~O. Dib, R.~Rosenfeld and A.~Zerwekh, \emph{{Double Higgs production and
  quadratic divergence cancellation in little Higgs models with T parity}},
  \href{http://dx.doi.org/10.1088/1126-6708/2006/05/074}{\emph{JHEP} {\bf 05}
  (2006) 074}, [\href{http://arxiv.org/abs/hep-ph/0509179}{{\tt
  hep-ph/0509179}}].

\bibitem{Dawson:2012mk}
S.~Dawson, E.~Furlan and I.~Lewis, \emph{{Unravelling an extended quark sector
  through multiple Higgs production?}},
  \href{http://dx.doi.org/10.1103/PhysRevD.87.014007}{\emph{Phys. Rev.} {\bf
  D87} (2013) 014007}, [\href{http://arxiv.org/abs/1210.6663}{{\tt
  1210.6663}}].

\bibitem{Dolan:2012ac}
M.~J. Dolan, C.~Englert and M.~Spannowsky, \emph{{New Physics in LHC Higgs
  boson pair production}},
  \href{http://dx.doi.org/10.1103/PhysRevD.87.055002}{\emph{Phys. Rev.} {\bf
  D87} (2013) 055002}, [\href{http://arxiv.org/abs/1210.8166}{{\tt
  1210.8166}}].

\bibitem{Grober:2016wmf}
R.~Grober, M.~Muhlleitner and M.~Spira, \emph{{Signs of Composite Higgs Pair
  Production at Next-to-Leading Order}},
  \href{http://dx.doi.org/10.1007/JHEP06(2016)080}{\emph{JHEP} {\bf 06} (2016)
  080}, [\href{http://arxiv.org/abs/1602.05851}{{\tt 1602.05851}}].

\bibitem{deLima:2014dta}
D.~E. Ferreira~de Lima, A.~Papaefstathiou and M.~Spannowsky, \emph{{Standard
  model Higgs boson pair production in the ( $ b\overline{b} $ )( $
  b\overline{b} $ ) final state}},
  \href{http://dx.doi.org/10.1007/JHEP08(2014)030}{\emph{JHEP} {\bf 08} (2014)
  030}, [\href{http://arxiv.org/abs/1404.7139}{{\tt 1404.7139}}].

\bibitem{Alwall:2014hca}
J.~Alwall, R.~Frederix, S.~Frixione, V.~Hirschi, F.~Maltoni, O.~Mattelaer
  et~al., \emph{{The automated computation of tree-level and next-to-leading
  order differential cross sections, and their matching to parton shower
  simulations}}, \href{http://dx.doi.org/10.1007/JHEP07(2014)079}{\emph{JHEP}
  {\bf 07} (2014) 079}, [\href{http://arxiv.org/abs/1405.0301}{{\tt
  1405.0301}}].

\bibitem{Christensen:2009jx}
N.~D. Christensen, P.~de~Aquino, C.~Degrande, C.~Duhr, B.~Fuks, M.~Herquet
  et~al., \emph{{A Comprehensive approach to new physics simulations}},
  \href{http://dx.doi.org/10.1140/epjc/s10052-011-1541-5}{\emph{Eur. Phys. J.
  C} {\bf 71} (2011) 1541}, [\href{http://arxiv.org/abs/0906.2474}{{\tt
  0906.2474}}].

\bibitem{Degrande:2011ua}
C.~Degrande, C.~Duhr, B.~Fuks, D.~Grellscheid, O.~Mattelaer and T.~Reiter,
  \emph{{UFO - The Universal FeynRules Output}},
  \href{http://dx.doi.org/10.1016/j.cpc.2012.01.022}{\emph{Comput. Phys.
  Commun.} {\bf 183} (2012) 1201--1214},
  [\href{http://arxiv.org/abs/1108.2040}{{\tt 1108.2040}}].

\bibitem{Alloul:2013bka}
A.~Alloul, N.~D. Christensen, C.~Degrande, C.~Duhr and B.~Fuks,
  \emph{{FeynRules 2.0 - A complete toolbox for tree-level phenomenology}},
  \href{http://dx.doi.org/10.1016/j.cpc.2014.04.012}{\emph{Comput. Phys.
  Commun.} {\bf 185} (2014) 2250--2300},
  [\href{http://arxiv.org/abs/1310.1921}{{\tt 1310.1921}}].

\bibitem{Degrande:2014vpa}
C.~Degrande, \emph{{Automatic evaluation of UV and R2 terms for beyond the
  Standard Model Lagrangians: a proof-of-principle}},
  \href{http://dx.doi.org/10.1016/j.cpc.2015.08.015}{\emph{Comput. Phys.
  Commun.} {\bf 197} (2015) 239--262},
  [\href{http://arxiv.org/abs/1406.3030}{{\tt 1406.3030}}].

\bibitem{Hahn:2000kx}
T.~Hahn, \emph{{Generating Feynman diagrams and amplitudes with FeynArts 3}},
  \href{http://dx.doi.org/10.1016/S0010-4655(01)00290-9}{\emph{Comput. Phys.
  Commun.} {\bf 140} (2001) 418--431},
  [\href{http://arxiv.org/abs/hep-ph/0012260}{{\tt hep-ph/0012260}}].

\bibitem{Hirschi:2011pa}
V.~Hirschi, R.~Frederix, S.~Frixione, M.~V. Garzelli, F.~Maltoni and R.~Pittau,
  \emph{{Automation of one-loop QCD corrections}},
  \href{http://dx.doi.org/10.1007/JHEP05(2011)044}{\emph{JHEP} {\bf 05} (2011)
  044}, [\href{http://arxiv.org/abs/1103.0621}{{\tt 1103.0621}}].

\bibitem{Frixione:1995ms}
S.~Frixione, Z.~Kunszt and A.~Signer, \emph{{Three jet cross-sections to
  next-to-leading order}},
  \href{http://dx.doi.org/10.1016/0550-3213(96)00110-1}{\emph{Nucl. Phys. B}
  {\bf 467} (1996) 399--442}, [\href{http://arxiv.org/abs/hep-ph/9512328}{{\tt
  hep-ph/9512328}}].

\bibitem{Frederix:2009yq}
R.~Frederix, S.~Frixione, F.~Maltoni and T.~Stelzer, \emph{{Automation of
  next-to-leading order computations in QCD: The FKS subtraction}},
  \href{http://dx.doi.org/10.1088/1126-6708/2009/10/003}{\emph{JHEP} {\bf 10}
  (2009) 003}, [\href{http://arxiv.org/abs/0908.4272}{{\tt 0908.4272}}].

\bibitem{Artoisenet:2012st}
P.~Artoisenet, R.~Frederix, O.~Mattelaer and R.~Rietkerk, \emph{{Automatic
  spin-entangled decays of heavy resonances in Monte Carlo simulations}},
  \href{http://dx.doi.org/10.1007/JHEP03(2013)015}{\emph{JHEP} {\bf 03} (2013)
  015}, [\href{http://arxiv.org/abs/1212.3460}{{\tt 1212.3460}}].

\bibitem{Alwall:2014bza}
J.~Alwall, C.~Duhr, B.~Fuks, O.~Mattelaer, D.~G. Öztürk and C.-H. Shen,
  \emph{{Computing decay rates for new physics theories with FeynRules and
  MadGraph5\_aMC@NLO}},
  \href{http://dx.doi.org/10.1016/j.cpc.2015.08.031}{\emph{Comput. Phys.
  Commun.} {\bf 197} (2015) 312--323},
  [\href{http://arxiv.org/abs/1402.1178}{{\tt 1402.1178}}].

\bibitem{Ball:2014uwa}
{\scshape NNPDF} collaboration, R.~D. Ball et~al., \emph{{Parton distributions
  for the LHC Run II}},
  \href{http://dx.doi.org/10.1007/JHEP04(2015)040}{\emph{JHEP} {\bf 04} (2015)
  040}, [\href{http://arxiv.org/abs/1410.8849}{{\tt 1410.8849}}].

\bibitem{Sjostrand:2014zea}
T.~Sjöstrand, S.~Ask, J.~R. Christiansen, R.~Corke, N.~Desai, P.~Ilten et~al.,
  \emph{{An Introduction to PYTHIA 8.2}},
  \href{http://dx.doi.org/10.1016/j.cpc.2015.01.024}{\emph{Comput. Phys.
  Commun.} {\bf 191} (2015) 159--177},
  [\href{http://arxiv.org/abs/1410.3012}{{\tt 1410.3012}}].

\bibitem{deFavereau:2013fsa}
{\scshape DELPHES 3} collaboration, J.~de~Favereau, C.~Delaere, P.~Demin,
  A.~Giammanco, V.~Lemaître, A.~Mertens et~al., \emph{{DELPHES 3, A modular
  framework for fast simulation of a generic collider experiment}},
  \href{http://dx.doi.org/10.1007/JHEP02(2014)057}{\emph{JHEP} {\bf 02} (2014)
  057}, [\href{http://arxiv.org/abs/1307.6346}{{\tt 1307.6346}}].

\bibitem{Cacciari:2011ma}
M.~Cacciari, G.~P. Salam and G.~Soyez, \emph{{FastJet User Manual}},
  \href{http://dx.doi.org/10.1140/epjc/s10052-012-1896-2}{\emph{Eur. Phys. J.
  C} {\bf 72} (2012) 1896}, [\href{http://arxiv.org/abs/1111.6097}{{\tt
  1111.6097}}].

\bibitem{Cacciari:2008gp}
M.~Cacciari, G.~P. Salam and G.~Soyez, \emph{{The Anti-k(t) jet clustering
  algorithm}},
  \href{http://dx.doi.org/10.1088/1126-6708/2008/04/063}{\emph{JHEP} {\bf 04}
  (2008) 063}, [\href{http://arxiv.org/abs/0802.1189}{{\tt 0802.1189}}].

\bibitem{Chatrchyan:2012jua}
{\scshape CMS} collaboration, S.~Chatrchyan et~al., \emph{{Identification of
  b-quark jets with the CMS experiment}},
  \href{http://dx.doi.org/10.1088/1748-0221/8/04/P04013}{\emph{JINST} {\bf 8}
  (2013) P04013}, [\href{http://arxiv.org/abs/1211.4462}{{\tt 1211.4462}}].

\bibitem{Thaler:2010tr}
J.~Thaler and K.~Van~Tilburg, \emph{{Identifying Boosted Objects with
  N-subjettiness}},
  \href{http://dx.doi.org/10.1007/JHEP03(2011)015}{\emph{JHEP} {\bf 03} (2011)
  015}, [\href{http://arxiv.org/abs/1011.2268}{{\tt 1011.2268}}].

\bibitem{Ellis:2009me}
S.~D. Ellis, C.~K. Vermilion and J.~R. Walsh, \emph{{Recombination Algorithms
  and Jet Substructure: Pruning as a Tool for Heavy Particle Searches}},
  \href{http://dx.doi.org/10.1103/PhysRevD.81.094023}{\emph{Phys. Rev. D} {\bf
  81} (2010) 094023}, [\href{http://arxiv.org/abs/0912.0033}{{\tt 0912.0033}}].

\bibitem{Khachatryan:2014vla}
{\scshape CMS} collaboration, V.~Khachatryan et~al., \emph{{Identification
  techniques for highly boosted W bosons that decay into hadrons}},
  \href{http://dx.doi.org/10.1007/JHEP12(2014)017}{\emph{JHEP} {\bf 12} (2014)
  017}, [\href{http://arxiv.org/abs/1410.4227}{{\tt 1410.4227}}].

\bibitem{CMS-PAS-PFT-09-001}
{\scshape CMS} collaboration, \emph{{Particle-Flow Event Reconstruction in CMS
  and Performance for Jets, Taus, and MET}},  Tech. Rep. CMS-PAS-PFT-09-001,
  CERN, 2009. Geneva, Apr, 2009.

\bibitem{ATLAS-CONF-2014-018}
{\scshape ATLAS} collaboration, \emph{{Tagging and suppression of pileup jets
  with the ATLAS detector}},  Tech. Rep. ATLAS-CONF-2014-018, CERN, Geneva,
  May, 2014.

\bibitem{Krohn:2009th}
D.~Krohn, J.~Thaler and L.-T. Wang, \emph{{Jet Trimming}},
  \href{http://dx.doi.org/10.1007/JHEP02(2010)084}{\emph{JHEP} {\bf 02} (2010)
  084}, [\href{http://arxiv.org/abs/0912.1342}{{\tt 0912.1342}}].

\bibitem{Goldberger:1999wh}
W.~D. Goldberger and M.~B. Wise, \emph{{Bulk fields in the Randall-Sundrum
  compactification scenario}},
  \href{http://dx.doi.org/10.1103/PhysRevD.60.107505}{\emph{Phys. Rev. D} {\bf
  60} (1999) 107505}, [\href{http://arxiv.org/abs/hep-ph/9907218}{{\tt
  hep-ph/9907218}}].

\bibitem{Oliveira:2014kla}
A.~Oliveira, \emph{{Gravity particles from Warped Extra Dimensions, predictions
  for LHC}},  \href{http://arxiv.org/abs/1404.0102}{{\tt 1404.0102}}.

\bibitem{Backovic:2014ega}
M.~Backovi\'c, T.~Flacke, J.~H. Kim and S.~J. Lee, \emph{{Boosted Event
  Topologies from TeV Scale Light Quark Composite Partners}},
  \href{http://dx.doi.org/10.1007/JHEP04(2015)082}{\emph{JHEP} {\bf 04} (2015)
  082}, [\href{http://arxiv.org/abs/1410.8131}{{\tt 1410.8131}}].

\end{thebibliography}\endgroup
\bibliographystyle{JHEP}

\end{document}